\documentclass[12pt]{article}

\usepackage{a4, graphicx, amsmath, amsthm }
\usepackage{color,ulem, psfrag}

\newcommand {\dn}[1] {\boldsymbol #1}
\newcommand {\mt}[1] {\mathrm #1}

\newcommand {\be} {\begin {equation} }
\newcommand {\ee} {\end {equation} }
\newcommand {\bmath} {\begin {displaymath} }
\newcommand {\emath} {\end {displaymath} }

\newcommand {\by} {\dn{y}}

\newcommand {\ident} { \mt{I}_{n} }
\newcommand {\identst} { \mt{I}_{\nstar} }
\newcommand {\nstar} { n^* }

\newcommand {\bx} { \mt{X} }

\newtheoremstyle{mytheoremstyle}
{3pt}
{3pt}
{\itshape}
{}
{\scshape}
{:}
{.5em}
{}
\theoremstyle{mytheoremstyle}

\usepackage{harvard} 
\newcommand{\citep}{\citeasnoun}  
\newcommand{\citea}{\citeaffixed}    

\begin{document}
\normalem

\title{\vspace*{-0.5in}
\textbf{Power-Conditional-Expected Priors:\\ Using $g$-priors with Random Imaginary Data for Variable Selection}}

\author{
D.~Fouskakis\thanks{D.~Fouskakis is with the Department of
Mathematics, National Technical University of Athens, Zografou
Campus, Athens 15780 Greece; email
\texttt{fouskakis@math.ntua.gr}}, \
and I.~Ntzoufras\thanks{I.~Ntzoufras is with the Department of
Statistics, Athens University of Economics and Business, 76
Patision Street, Athens 10434 Greece; email
\texttt{ntzoufras@aueb.gr}}}

\date{}

\maketitle

\noindent
\textbf{Summary:}

The Zellner's $g$-prior and its recent hierarchical extensions are the most popular default prior choices 
in the Bayesian variable selection context. 
These prior set-ups can be expressed power-priors with fixed set of imaginary data. 
In this paper, we borrow ideas from the power-expected-posterior (PEP) priors 
in order to introduce, under the $g$-prior approach, an extra hierarchical level that accounts for the imaginary data uncertainty. 
For normal regression variable selection problems, the resulting power-conditional-expected-posterior (PCEP) prior is a conjugate normal-inverse gamma prior which provides a consistent variable selection procedure and gives support to more parsimonious models than the ones supported using the $g$-prior and the hyper-$g$ prior for finite samples.
Detailed illustrations and comparisons of the variable selection procedures using the proposed method, 
the $g$-prior and the hyper-$g$ prior are provided using both simulated and real data examples. 

\vspace*{0.15in}


\section{Introduction}

During the last years, research in Bayesian variable selection has been focused on the choice of suitable and meaningful priors
for the model parameters.  Specification of the
hyperparameters of proper priors for model-specific parameters is crucial since posterior model odds are highly sensitive
on the values of the prior variances due to the Jeffreys-Lindley-Bartlett paradox \cite{lindley_57,bartlett_57}. Moreover, the use of improper
priors is not allowed, due to the presence of unknown normalizing constants
involved in the calculation of the Bayes factors.
A series of important
publications deal with the above mentioned issues, including
the $g$-prior \cite{zellner_86},
the benchmark priors of \cite{fernandez_etal_2001},
the fractional Bayes factor approach \cite{ohagan_95},
the intrinsic Bayes factor \cite{berger_pericchi_96b},
the intrinsic variable selection method \cite{casella_moreno_2006}
 and the expected-posterior prior approach \cite{perez_berger_2002} among others.
More recently, interest lies on the mixtures of $g$-priors, including the hyper-$g$ prior of
\citep{liang_etal_2008}, the extension of \citep{bove_held_2011} for GLMs and the work of \citep{ley_steele_2012} 
for economic applications.

A usual mechanism to produce sensible and compatible prior distributions across models is via imaginary data.
The Zellner's $g$-prior can be expressed as a power-prior with fixed set of imaginary data; see for details \citep{zellner_86}
and \citep{ibrahim_chen_2000}. Similar is the case for any mixture of $g$-prior, with additional uncertainty introduced on
the volume of the information  that the imaginary data account in the posterior inference.

In this article, we further use ideas from the expected-posterior prior approach \cite{perez_berger_2002}
in order to introduce uncertainty around the assumed imaginary data in a similar manner as in \citep{fouskakis_et.al_2013}.
Specifically, we introduce a hyperprior for the imaginary data by adding
an extra hierarchical level to our model structure that has an effect on the prior mean of the regression coefficients.

When our approach is implemented in the Zellner's $g$-prior, the result
is a normal-inverse gamma conjugate prior that leads to a variable selection procedure that is similar, for large datasets, but systematically more parsimonious, for small sample sizes, than the one using the Zellner's $g$-prior 
or mixtures  of $g$-priors.

The plan of the paper is as follows. In Section \ref{section2} we discuss the role of imaginary data in $g$-priors. In Section \ref{section3} we discuss the extension of the $g$-prior by considering imaginary data coming from a ``suitable" predictive distribution using the expected-posterior prior approach. 
Our new prior and the induced variable selection procedure, under a specific choice of baseline prior, is fully described in detail in Section \ref{section4}; formulas for the resulting prior, posterior and marginal likelihood are given and a short discussion about the choice of hyperparameters is presented. Section \ref{consistency} explores the limiting behaviour of the marginal likelihood, while in Section \ref{interpretation} we discuss the differences between our prior and the Zellner's $g$-prior. In Section \ref{examples} we present illustrations of our method and Section \ref{sec_discussion} concludes the paper with a brief discussion.  

\section{The role of imaginary data in $g$-priors}
\label{section2}
Let us consider a set of imaginary data $\by^* = ( y_1^*,y_2^*, \dots, y_{n^*}^*  )^T$ of size $n^*$.
Then, following the power-prior approach introduced by \citep{ibrahim_chen_2000}, for any model $m_\ell$ with parameter vector $\dn{\theta}_\ell$, likelihood $f(\by^*| \dn{\theta}_\ell, m_\ell)$
and baseline prior $\pi^N_\ell( \dn{\theta}_\ell)$,
we can obtain a ``sensible'' prior for the model parameters based on the following expression
$$
\pi_\ell( \dn{\theta}_\ell | \by^*; \delta ) \propto f(\by^*| \dn{\theta}_\ell, m_\ell)^{1/\delta} \pi^N_\ell( \dn{\theta}_\ell)\,.
$$
The parameter $\delta \geq 1$ controls the weight that the imaginary data contribute to the ``final'' posterior distribution of  $\dn{\theta}_\ell$, since
$$
\pi_\ell( \dn{\theta}_\ell | \by, \by^*; \delta ) \propto f(\by| \dn{\theta}_\ell, m_\ell)f(\by^*| \dn{\theta}_\ell, m_\ell)^{1/\delta} \pi^N_\ell( \dn{\theta}_\ell)\,.
$$
For $\delta=1$, the above prior is exactly equal to the posterior distribution of $\dn{\theta}_\ell$ after
observing the imaginary data $\by^*$. For $\delta=1/n^*$ the contribution of the imaginary data to the overall posterior
is equal to one data point; i.e. the prior has a unit-information interpretation \cite{kass_wasserman_95}.

In the following we focus on variable selection problems for normal regression models.
Therefore, for any model $m_\ell$, with parameters $\dn{\theta}_\ell = ( \dn{\beta}_\ell\,, \sigma^2 )$
the likelihood is specified by
\begin{equation}
\label{nm}
\dn{Y} | \mt{X}_\ell, \dn{\beta}_\ell, \sigma^2, m_\ell \sim N_n (
\mt{X}_\ell \, \dn{\beta}_\ell \,, \sigma^2 \mt{I}_n )
\end{equation}
where
$\dn{Y}=(Y_1, \dots, Y_n)^T$ is a multivariate random variable expressing the response for each subject,
$\mt{X}_\ell$ is a $n \times d_\ell$ design matrix containing the values of the explanatory variables in its columns,
$\mt{I}_n$ is the $n\times n$ identity matrix,
$\dn{\beta}_\ell$ is a vector of length $d_\ell$ with the effects of each covariate on the response variable and
$\sigma^2$ is the error variance, common to all models.

If we adopt the power-prior approach for the regression coefficients $\dn{\beta}_\ell$ given $\sigma^2$, with imaginary data $\dn{y}^*$, of size $n^*$
and imaginary design matrix $\mt{X}_\ell^*$, then the prior will be defined as
$$
\pi_\ell( \dn{\beta}_\ell | \sigma^2, \by^*; \delta )
\propto
\exp \left(  -\frac{1}{2 \delta \sigma^2} \left( \by^*- \mt{X}_\ell^* \, \dn{\beta}_\ell \right)^T \left( \by^*- \mt{X}_\ell^* \, \dn{\beta}_\ell \right) \right) \pi^N_\ell( \dn{\beta}_\ell|\sigma^2),
$$
with $\pi^N_\ell( \dn{\beta}_\ell|\sigma^2)$ denoting the baseline prior for $\dn{\beta}_\ell$ given $\sigma^2$.
When $\pi^N_\ell( \dn{\beta}_\ell|\sigma^2) \propto 1$, then
$$
\pi_\ell( \dn{\beta}_\ell | \sigma^2, \by^*; \delta )
=
f_{N_{d_\ell}} \Big( \dn{\beta}_\ell \, ; \, \widehat{\dn{\beta}}_\ell^*\, , \delta \big( \bx_\ell^{*T}\bx_\ell^* \big)^{-1} \sigma^2 \Big),
$$
where
$\widehat{\dn{\beta}}_\ell^* = \big( \bx_\ell^{*T}\bx_\ell^* \big)^{-1} \bx_\ell^{*T} \by^*$ and
$f_{N_d}( \by \, ; \, \dn{\mu}, \dn{\Sigma} )$ denoting the density of the d-dimensional normal distribution
with mean $\dn{\mu}$ and variance-covariance matrix $\dn{\Sigma}$ evaluated at $\by$.
>From the above, it is obvious that the Zellner's $g$-prior can be expressed as a power-prior
using imaginary data with the same design matrix as the original,
i.e. $\bx_\ell^* = \bx_\ell$, mean equal to
$\dn{\mu} =  \big( \bx_\ell^{T}\bx_\ell \big)^{-1} \bx_\ell^{T} \by^*$ and $g=\delta$.
The usual case with zero mean is simply obtained assuming imaginary data $\dn{y}^*=\dn{0}$,
i.e. the imaginary data are coming from the constant model with zero mean and no variability.

A similar expression is obtained even if the baseline prior has the following $g$-prior structure:
\be
\label{prior_cond}
\pi^N_\ell( \dn{\beta}_\ell|\sigma^2) =
f_{N_{d_\ell}} \Big( \dn{\beta}_\ell \, ; \, \textbf{0},  g_0 \big( \bx_\ell^{*T}\bx_\ell^* \big)^{-1} \sigma^2 \Big).
\ee
In this case, the power-prior is given by 
\be
\pi_\ell(\dn{\beta}_\ell\, | \sigma^2, \by^*; \delta) =
f_{N_{d_\ell}} \big(\dn{\beta}_\ell \,;\, w \widehat{\dn{\beta}}_\ell^*, w \delta ( \mt{X}_\ell^{*^T} \mt{X}_\ell^*)^{-1} \sigma^2\big),
\label{bgp}
\ee
where $w=g_0/(g_0+\delta)$. Thus, the Zellner's $g$-prior can be interpreted now as a power-prior
using imaginary data with the same design matrix as the original,
i.e. $\bx_\ell^* = \bx_\ell$, mean equal to
$\dn{\mu} =  w \big( \bx_\ell^{T}\bx_\ell \big)^{-1} \bx_\ell^{T} \by^*$ and $g=w\delta$.

Furthermore, the modified version of the $g$-prior, as in \citep{liang_etal_2008}:
\be
\label{mod_g}
\pi_\ell( \dn{\beta}_{\setminus 0, \ell} | \beta_{0, \ell}, \sigma^2, \by^* )
=
f_{N_{d_\ell}} \Big( \dn{\beta}_{\setminus 0, \ell} \, ; \, \dn{0} ,
g \big( \bx_{\setminus 0, \ell}^{T}\bx_{\setminus 0, \ell} \big)^{-1} \sigma^2 \Big),
\ee
with $\dn{\beta}_{\setminus 0, \ell}\,$ denoting the sub-vector of $\dn{\beta}_{\ell}$ without the constant parameter
$\beta_{0, \ell}$, can be also interpreted as a power-prior using similar arguments as above.
Specifically, we can obtain (\ref{mod_g}) by assuming imaginary data
$\dn{y}^* = \beta_{0, \ell}\, \dn{1}_{n}$ for a given $\beta_{0, \ell}$ since the linear predictor of the regression model is written
as
$\mt{X}_\ell \, \dn{\beta}_\ell = \beta_{0, \ell} \dn{1}_{n} + \mt{X}_{\setminus 0, \ell} \, \dn{\beta}_{\setminus 0, \ell}$;
where $\dn{1}_{n}$ is the vector of length $n$ with all elements equal to one.

\section{Random imaginary data and $g$-priors}
\label{section3}

The hyperparameter $g$ in the $g$-prior, controls the inverse relative imaginary sample size. Over the last years reasearch has been focused on the selection
of this hyperparameter \citeaffixed{george_forster_2000,fernandez_etal_2001}{e.g.}. Lately, \citep{liang_etal_2008} studied mixtures of $g$-priors by
introducing the use of an hyperprior for $g$. In all the developments of the $g$-prior the imaginary data are assumed to be fixed as described in the previous section.
Here we extend the $g$-prior (with $g$ either fixed or random) in a different direction, by considering imaginary data coming from
a ``suitable" predictive distribution. Specifically, we add an extra hierarchical level to the specification of the prior distribution, that
has an effect on both the prior mean and the prior variance, through the variability of the imaginary data. Therefore,
for any model $m_\ell$, the resulting prior for $\dn{\beta}_\ell$, given $\sigma^2$
has the following form
\be
\label{hgp}
\pi_\ell( \dn{\beta}_\ell | \sigma^2, \by^*)=\int f_{N_{d_\ell}} \Big( \dn{\beta}_\ell \, ; \, w\widehat{\dn{\beta}}_\ell^{*z}\, ,
g \big( \bx_\ell^{T}\bx_\ell \big)^{-1} \sigma^2 \Big) m^*(\by^*) d\by^*
\ee
where
$\widehat{\dn{\beta}}_\ell^{*z} = \big( \bx_\ell^{T}\bx_\ell \big)^{-1} \bx_\ell^{T} \by^*$ and $m^*(\by^*)$ is the hyperprior for the imaginary data
$\by^*$ of size $n$. In the above expression $w$ and $g$ are hyperparameters that need to be specified; details are shown later in this Section.

For the specification of the hyperprior $m^*$, we might use the expected-posterior prior approach \cite{perez_berger_2002},
that, as will see in the next section, assumes random imaginary samples coming from a common underlying predictive distribution,
using an initial baseline prior distribution.

\subsection{Expected-posterior priors}

\citep{perez_berger_2002} have defined the expected-posterior (EP) prior
as the posterior distribution of a parameter vector of the model under consideration
averaged over all possible samples $\by^*$ coming from the predictive distribution $f(\by^*|m_0)$ of a reference model $m_0$
\cite[def. 1, p. 493]{perez_berger_2002}.
Hence the EP prior for the parameter vector $\dn{\theta}_\ell = ( \dn{\beta}_\ell\,, \sigma^2 )$
of any model $m_\ell \in \cal{M}$, where $\cal M$ is the model space, is given by
\be
\begin{split}
\pi_\ell^{EP}( \dn{\theta}_\ell )
& =  \int \pi_\ell^N ( \dn{\theta}_\ell | \by^*) m_0^N(\by^*) d\by^* ,
\end{split}
\label{epp}
\ee
where $\pi_\ell^N ( \dn{\theta}_\ell | \by^*)$ is the posterior of $\dn{\theta}_\ell$ for model $m_\ell$
using a baseline prior $\pi_\ell^N(\dn{\theta}_\ell)$ and
$m_0^N(\by^*)$ is the prior predictive distribution, evaluated at $\by^*$, for model $m_0$
under the prior $\pi_0^N(\dn{\theta}_0)$.

\subsection{Power-conditional-expected-posterior (PCEP) priors}
\label{sec_PCEP}

Since $\sigma^2$ appears in all models under comparison, we can assume a common prior distribution $\pi^{N}_\ell( \sigma^2 )$
for all models $m_\ell \in {\cal M}$. This is a usual practice in the related literature as noted by \citep{fernandez_etal_2001}
and references therein. Hence, we may implement the EP prior approach only for the regression coefficients $\dn{\beta}_\ell$
conditionally on the values of the error variance $\sigma^2$. By this way, we define the conditional-expected-posterior (CEP) prior  by
$$
\pi^{CEP}_\ell ( \dn{\beta}_\ell, \sigma^2_\ell ) = \pi^{CEP}_\ell ( \dn{\beta}_\ell | \sigma^2_\ell ) \pi^{N}_\ell( \sigma^2 )
$$
with
\begin{equation*}
\pi^{CEP}_\ell ( \dn{\beta}_\ell | \sigma^2_\ell )
 =   \int \pi_\ell^N ( \dn{\beta}_\ell | \sigma^2, \by^*) m_0^N(\by^* | \sigma^2 ) d\by^* .
\end{equation*}

Furthermore, in order to diminish the effect of the training samples, we use ideas from the power-expected-posterior prior approach as defined by
\citep{fouskakis_et.al_2013}. Thus we introduce the \textit{power-conditional-expected-posterior} (PCEP) prior by raising
the likelihood, involved in the CEP prior, to a power $1/\delta$ that controls the effect of the training sample in the PCEP prior.
Therefore, the PCEP prior is defined as
\begin{equation}
\label{pcep}
\pi^{PCEP}_\ell ( \dn{\beta}_\ell, \sigma^2_\ell; \delta ) = \pi^{PCEP}_\ell ( \dn{\beta}_\ell | \sigma^2_\ell ; \delta) \pi^{N}_\ell( \sigma^2 )
 =   \left[\int \pi_\ell^N ( \dn{\beta}_\ell | \sigma^2, \by^*; \delta) m_0^N(\by^* | \sigma^2; \delta ) d\by^* \right] \pi^{N}_\ell( \sigma^2 ),
\end{equation}
where
$$
\pi_\ell^N ( \dn{\beta}_\ell | \sigma^2, \by^*; \delta) = \frac{f( \by^* | \, \dn{\beta}_\ell\,,  \sigma^2, m_\ell \, ; \mt{X}^*_\ell \, , \delta )
\pi_\ell^{N}(\dn{\beta}_\ell\,|\,\sigma^2;\mt{X}_\ell^*)}{m_\ell^N (\by^*|\,\sigma^2;\, \mt{X}_\ell^*\, , \delta)}
$$
with $f( \by^* | \, \dn{\beta}_\ell\,,  \sigma^2, m_\ell \, ; \mt{X}_\ell^* \, , \delta )
\propto  f( \by^* | \dn{\beta}_\ell\,, \sigma^2, m_\ell \, ; \mt{X}_\ell^*)^{1/\delta}$
 being the density-normalized power-likelihood given, in our case, by
\begin{eqnarray}
\label{pl}
f( \by^* | \, \dn{\beta}_\ell\,,  \sigma^2, m_\ell \, ; \mt{X}_\ell^* \, , \delta )
 &=&  f_{N_{n^*}}(\by^* \,;\, \mt{X}_\ell^* \dn{\beta}_\ell\,, \delta \sigma^2 \mt{I}_{n^*})~ \label{power_likelihood}.
\end{eqnarray}
Moreover, $m_\ell^N( \by^*|\sigma^2;\, \mt{X}_\ell^*\, , \delta)$
is the prior predictive distribution (or the marginal likelihood), evaluated at $\by^*$, of model $m_\ell$ given $\sigma^2$
with the power-likelihood defined by (\ref{power_likelihood})
under the baseline prior $\pi^N_\ell( \dn{\beta}_\ell\,| \,\sigma^2 ;\mt{X}_\ell^*)$, i.e.
\begin{eqnarray*}
m_\ell^N(\by^*|\,\sigma^2;\, \mt{X}_\ell^*\, , \delta)
 &=& \int f_{N_{n^*}}(\by^* \,;\, \mt{X}_\ell^* \dn{\beta}_\ell\,, \delta \sigma^2 \mt{I}_{n^*})
        \pi^N_\ell( \dn{\beta}_\ell\,| \, \sigma^2;\mt{X}_\ell^*) d\dn{\beta}_\ell~.
\end{eqnarray*}

As discussed in \citep{fouskakis_et.al_2013}, we can set the power-parameter $\delta$ equal to $n^*$, to represent prior information equal to one data point. In a similar manner as in the $g$-prior, we set $n^*=n$ (and therefore $\mt{X}_\ell^* = \mt{X}_\ell$); by this way we also dispense with the selection of the training samples. 

\subsection{Using PCEP prior for the specification of the hyperprior for the imaginary data}
\label{section3.3}
As we have already seen in Section \ref{section2}, the posterior $\pi_\ell^N ( \dn{\beta}_\ell | \sigma^2, \by^*; \delta)$ involved in the
definition of the PCEP prior takes the form
$f_{N_{d_\ell}} \big(\dn{\beta}_\ell \,;\, w \widehat{\dn{\beta}}_\ell^*, w \delta ( \mt{X}_\ell^{*^T} \mt{X}_\ell^*)^{-1} \sigma^2\big)$, with
$w=g_0/(g_0+\delta)$,
when the baseline prior of $\dn{\beta}_\ell$ given $\sigma^2$ is (\ref{prior_cond}).

Thus (\ref{hgp}) can be obtained as the PCEP prior (\ref{pcep}) with
$g=w\delta$, $\mt{X}_\ell^* = \mt{X}_\ell$ and the hyperprior for the imaginary data of size $n^*=n$ given by $m^*(\by^*)=m_0^N(\by^* | \sigma^2; \delta )$,
i.e. the prior predictive of the reference model, evaluated using the power-likelihood (\ref{power_likelihood}) and the baseline prior (\ref{prior_cond}).

A question which naturally arises is which model must be selected as a reference model.
\citep{perez_berger_2002} indirectly supported the choice of the most parsimonious model in ${\cal M}$.
This choice provides a sensible interpretation since we a-priori argue
in favor of the assumption that the data are coming from the simplest model
supporting by this way the parsimony principle. The latter interpretation is close to the sceptical prior approach
as described by \citep[Section 5.5.2] {spiegelhalter_abrams_myles}
where a tendency toward the null hypothesis must be a-priori supported
by centering our prior beliefs around values assumed by this hypothesis
when no other information is available. 
The constant model (with no predictors) can naturally serve as the reference model in our case.

\section{PCEP $g$-prior methodology}
\label{section4}

In this section we implement the PCEP prior introduced in Section \ref{sec_PCEP} using
the Zellner's $g$-prior (\ref{prior_cond}) as baseline.
Furthermore, we assume an $IG(a,b)$ prior distribution for $\sigma^2$.
Then, for any model $m_\ell\,$, the prior predictive distribution, under the baseline prior, conditional on $\sigma^2$,
is  a multivariate normal distribution given by
\be
m_\ell^N(\by^*|\sigma^2;\, \mt{X}_\ell^*\, , \delta) = f_{N_{n}} \big( \by^* \, ; \, \dn{0},   {\Lambda_\ell^*}^{-1} \sigma^2 \big)~,
\label{prior_predictive_cond}
\ee
where
\begin{equation}
{\Lambda_\ell^*}^{-1}
= \delta \Big( \mt{I}_{n^*}-\frac{g_0}{g_0+\delta} \mt{X}_\ell^*\left({\mt{X}_\ell^*}^T\mt{X}_\ell^*\right)^{-1}{\mt{X}_\ell^*}^T \Big)^{-1}
= \delta \mt{I}_{n^*} + g_0 \mt{X}_\ell^*\left({\mt{X}_\ell^*}^T\mt{X}_\ell^*\right)^{-1}{\mt{X}_\ell^*}^T ~.
\label{lambda}
\end{equation}
Derivation of the above marginal likelihood is given in Appendix \ref{proof_prior_predictive_cond}. For the special case of the constant model, the
variance--covariance matrix of the above distribution simplifies
to $\big[ \delta  \mt{I}_{n^*}+ g_0
{n^*}^{-1}\dn{1}_{n^*} \dn{1}_{n^*}^T
 \big]\sigma^2$; where $\dn{1}_{n^*}$ is a vector of length $n^*$
with all elements equal to one.

\subsection{Prior distribution}
\label{power_conditional_intrinsic_prior}

The power-conditional-expected-posterior (PCEP) prior on $\dn{\beta}_\ell$ given $\sigma^2$
is
\small
\begin{eqnarray}
\pi_\ell^{PCEP}(\dn{\beta}_\ell\, , \sigma^2 | \mt{X}_\ell^*\, ,\delta)&=&\pi_\ell^{PCEP}(\dn{\beta}_\ell \; |\sigma^2 ; \mt{X}_\ell^*\, ,\delta)\pi_\ell^{N}(\sigma^2)\\
&=& \left[
    \int
    \frac{
            f(\by^*  |  \dn{\beta}_\ell\,, \sigma^2, m_\ell\,;\bx_\ell^*\,,\delta)
            \pi^N_\ell ( \dn{\beta}_\ell \;| \sigma^2 ; \mt{X}_\ell^*)}
         {m_\ell^N (\by^* | \sigma^2;\,\mt{X}_\ell^*\, ,\delta)}
          m_0^N(\by^* | \sigma^2;\,\mt{X}_0^*\, ,\delta) d\by^* \right]
f_{IG}\left(\sigma^2 \,;\,a , b  \right) \nonumber
\\
&=&
\left[
\int
 \pi_\ell^{N}(\dn{\beta}_\ell\, | \by^*, \sigma^2 \,; \mt{X}_\ell^*\, ,\delta) m_0^N(\by^* | \sigma^2;\,\mt{X}_0^*\, ,\delta) d\by^* \right]
f_{IG}\left(\sigma^2 \,;a , b \right) \nonumber
\\
&=&
\left[
\int
f_{N_{d_\ell}} \big(\dn{\beta}_\ell \,;\, w \widehat{\dn{\beta}}_\ell^*, w \delta ( \mt{X}_\ell^{*^T} \mt{X}_\ell^*)^{-1} \sigma^2\big)
f_{N_{n^*}} \big( \by^* \, ; \, \dn{0},   \Lambda_0^{*^{-1}} \sigma^2 \big) d\by^* \right]
f_{IG}\left(\sigma^2 \,;a , b  \right) \nonumber
\\
&=& f_{N_{d_\ell}} \Big(\dn{\beta}_\ell \,;\, \dn{0}, \; \delta
\left\{ \bx_\ell^{*^T} \left[ w^{-1}  \identst - ( \delta
\Lambda_0^{*}  + w \mt{H}_\ell^* )^{-1} \right] \bx_\ell^*
\right\}^{-1} \hspace{-1ex} \sigma^2 \Big) f_{IG}\left(\sigma^2
\,;a , b  \right),
\label{PCEP_prior}
\end{eqnarray}
\normalsize
where $f_{IG}\left( y \, ; \, a, b \right)$ denotes the density of the
inverse gamma distribution with parameters $a$ and $b$ and mean
equal to $b/(a-1)$ evaluated at $y$. Additionally, $\pi^N_\ell(\dn{\beta}_\ell\, | \by^*, \sigma^2 \,; \mt{X}_\ell^*\, ,\delta)$ can be considered as
a conditional posterior of $\dn{\beta}_\ell\, | \sigma^2$ with power-likelihood (\ref{power_likelihood})
and prior (\ref{prior_cond}) and is given by (\ref{bgp}); details are provided
in the Appendix \ref{proof_cond_posterior}.  Furthermore $\mt{H}_\ell^* = \mt{X}_\ell^{*}(\mt{X}_\ell^{*^T}\mt{X}_\ell^*)^{-1}\mt{X}_\ell^{*^T}$.

\subsection{Posterior distribution}
\label{sec_posterior_PCEP}

The above resulting prior is the usual conjugate normal--inverse
gamma prior with mean equal to $\dn{0}$, scale parameter equal to
\be
\mt{V_{\dn{\beta}_\ell}^*}= \delta \left\{ \bx_\ell^{*^T} \left[
w^{-1} \identst - ( \delta \Lambda_0^{*} + w \mt{H}_\ell^* )^{-1}
\right] \bx_\ell^* \right\}^{-1}
\label{v_star}
\ee
and parameters $a$ and $b$ for the inverse--gamma component. 
Hence, the posterior distribution under the
power-conditional-expected-posterior (PCEP) prior on $\dn{\beta}_\ell$ given $\sigma^2$
is a normal inverse gamma distribution, i.e.
\begin{eqnarray*}
\pi_\ell^{PCEP}(\dn{\beta}_\ell, \sigma^2 | \by; \mt{X}_\ell\,, \mt{X}_\ell^*,
\; \delta ) &=& f_{N_{d_\ell}} \big(\dn{\beta}_\ell \,;\,
\widetilde{\dn{\beta}}, \; \widetilde{\mt{\Sigma}} \sigma^2 \big)
f_{IG}\big(\sigma^2 \,;\, \widetilde{a}_\ell , \widetilde{b}_\ell
\big),
\end{eqnarray*}
where
$$
\widetilde{\dn{\beta}}  = \widetilde{\mt{\Sigma}} \mt{X}_\ell^T
\by, ~ \widetilde{\mt{\Sigma}} =
\left\{\mt{V_{\dn{\beta}_\ell}^*}{^{-1}}+ \mt{X}_\ell^T
\mt{X}_\ell \right\}^{-1}, \widetilde{a}_\ell = n/2 + a, ~
\widetilde{b}_\ell = SS_\ell/2 + b
$$
with
$SS_\ell
= \by^T \big(  \ident - \mt{X}_\ell^T \widetilde{\mt{\Sigma}}\mt{X}_\ell \big)  \by
= \by^T \big(  \ident + \mt{X}_\ell^T \mt{V_{\dn{\beta}_\ell}^*} \mt{X}_\ell \big)^{-1} \by$.

\subsection{Marginal likelihood}
\label{sec_marginal_likelihood_PCEP}

The marginal likelihood, under the PCEP $g$-prior approach is given by
\be
m^{PCEP}_\ell( \by | \mt{X}_\ell\,,\mt{X}_\ell^*\, ,\delta) = f_{St_n} \left( \by \, ; ~
2a, ~\dn{0}, ~ \frac{b}{a} \Big[ \mt{I}_n +
\mt{X}_\ell \mt{V_{\dn{\beta}_\ell}^*} \mt{X}_\ell^T \Big]
\right),
\label{marginal_likelihood_power_cond_intrinsic}
\ee
in which $f_{St_n} ( \by \, ; d, \dn{ \mu }, \Sigma )$ is the density of the multivariate
Student distribution in $n$ dimensions with $d$ degrees of freedom,
location $\dn{ \mu }$ and scale $\Sigma$.

Since the above marginal likelihood can be calculated analytically,
we can directly compare all models without any problem and identify
the maximum a-posteriori (MAP) model,
the median probability (MP) model or
the best equally well behaved models with Bayes factors less than 3 when compared with the MAP
according to the interpretation table of \citep{kass_raftery_95}.

When the model space is large we can implement $MC^3$ \cite{madigan_york_95}
to explore the model space and trace the best models (see Appendix \ref{sec_model_search}).

\subsection{Specification of prior parameters}
\label{section4.4}
Clearly the marginal likelihood for the PCEP methodology depends on the selection of the power parameter $\delta$, the training sample
and its size $n^*$, the reference model $m_0$ and the prior hyperparameters $g_0$, $a$ and $b$.
Following \citep{fouskakis_et.al_2013} we propose
\begin{itemize}

\item the power parameter $\delta$ to be equal to $n^*$ in order to account the data for information equal to one data point. If additionally we set
$n^*=n$, and therefore $\mt{X}_\ell^* = \mt{X}_\ell$, we avoid completely the training sample and its possible effect to the posterior model
comparison inference, while we account still for information equal to one data point.

\item the parameter $g_0$ in the normal baseline prior is set equal to $\delta n^*$.
            Therefore, for $\delta=n^*$ we propose to use $g_0=n^{*2}$.
            This choice will make the baseline Zellner's $g$-prior to contribute with information equal to one data point
            within the posterior
            $\pi_\ell^{N}( \dn{\beta}_\ell\,  |  \sigma^2, \,\by^*\, ; \mt{X}_\ell^* \, , \delta )$.
            By this way, the whole PCEP prior will account to information equal to $1+1/\delta$ data points.
            \color{black}

\item the parameters $a$ and $b$ in the inverse gamma baseline prior to be equal to 0.01 in order to have a baseline prior mean 1 and variance equal to 100 (i.e. large)
for the precision parameter.

\item the reference model $m_0$ to be the constant model as discussed in Section \ref{section3.3}. With this choice we also avoid
the need for the specification of the imaginary design matrix, since $\mt{X}_0^* = \dn{1}_{n^*}$.

\item the size of the training sample $n^*$ to be $n$.

\end{itemize}

\section{Limiting behaviour of the marginal likelihood}
\label{consistency}

From (\ref{marginal_likelihood_power_cond_intrinsic}), we have that
$$
\log m^{PCEP}_\ell( \by | \mt{X}_\ell\,,\mt{X}_\ell^*\, ,\delta)
= C - \frac{1}{2} \log | \mt{I}_n + \mt{X}_\ell \mt{V_{\dn{\beta}_\ell}^*} \mt{X}_\ell^T |
- \left( \frac{n}{2} + a \right)
\log \left( 2b  + \by^T \big( \mt{I}_n + \mt{X}_\ell \mt{V_{\dn{\beta}_\ell}^*} \mt{X}_\ell^T \big)^{-1} \by \right), ~
$$
where $C$ is a constant that does not depends on the model structure $m_\ell$. We set $n^*=n$, $\mt{X}_\ell^* = \mt{X}_\ell$ and let
$\mt{V_{\dn{\beta}_\ell}}$, $\mt{H}_\ell$ and $\mt{\Lambda}_0$ defined as $\mt{V_{\dn{\beta}_\ell}^*}$, $\mt{H}_\ell^*$ and $\mt{\Lambda}_0^{*}$ by replacing
$\mt{X}_\ell^* = \mt{X}_\ell$.

The determinant involved in the above expression is equal to
\be
| \mt{I}_n + \mt{X}_\ell \mt{V_{\dn{\beta}_\ell}} \mt{X}_\ell^T | =
\left( 1 + \delta w \right)^{d_\ell}
\left|   \Lambda_0 \right|^{-1}
\left|    \Lambda_0 +  \left( \frac{     w^2  }{ 1+\delta w }\right) \mt{H}_\ell   \right|
\label{consistency1}
\ee
while
\be
\by^T \big( \mt{I}_n + \mt{X}_\ell \mt{V_{\dn{\beta}_\ell}} \mt{X}_\ell^T \big) \by
=\by^T\by - \frac{1+w \delta }{ w \delta }  \by^T\mt{X}_\ell
  \left(
   \bx_\ell^{T}\bx_\ell - \frac{ w  }{1+w \delta }
   \bx_\ell^{T}( \mt{I}_{n}  + w [\mt{H}_\ell-\mt{H}_0])^{-1}\bx_\ell\right)^{-1}  \mt{X}_\ell^T \by;   \label{consistency2}
\ee
for detailed derivations of these two identities see Appendix \ref{appendix_calc_concistency}.

For large $n$ and for the proposed hyperparameter values (see Section (\ref{section4.4})) we obtain
\be
| \mt{I}_n + \mt{X}_\ell \mt{V_{\dn{\beta}_\ell}} \mt{X}_\ell^T | =
(nw+1)^{d_\ell} \frac{ |  \mt{\Lambda}_0 + \frac{w^2}{nw+1}P_\ell| }{ |   \mt{\Lambda}_0 | }
\approx (n+1)^{d_\ell},
\label{consistency3}
\ee
while
\begin{eqnarray}
\by^T \big( \mt{I}_n + \mt{X}_\ell \mt{V_{\dn{\beta}_\ell}} \mt{X}_\ell^T \big) \by
 &\approx&
\by^T\by - \frac{1+  \delta }{   \delta }  \by^T\mt{X}_\ell
  \left(
   \bx_\ell^{T}\bx_\ell - \frac{  1  }{1+  \delta }
   \bx_\ell^{T}( \mt{I}_{n}  +   [\mt{H}_\ell-\mt{H}_0])^{-1}\bx_\ell\right)^{-1}  \mt{X}_\ell^T \by  \hspace{3em} \nonumber \\
&\approx&
\by^T\by -   \by^T\mt{X}_\ell
  \left(
   \bx_\ell^T\bx_\ell \right)^{-1}  \mt{X}_\ell^T \by  \equiv RSS_\ell.
\end{eqnarray}

Therefore, the log marginal likelihood can be approximated by
\begin{eqnarray*}
\log m^{PCEP}_\ell( \by | \mt{X}_\ell\,,\mt{X}_\ell^*\, ,\delta)
&\approx&
C - \frac{d_\ell}{2} \log (n+1) - \left( \frac{n}{2} + a \right)  \log \left( 2b  + RSS_\ell \right)\\
&\approx&
C - \frac{d_\ell}{2} \log (n) -  \frac{n}{2}   \log RSS_\ell \\
&\approx&
C - \frac{1}{2} BIC_\ell~.
\end{eqnarray*}

Hence, PCEP $g$-prior has the same limiting behavior as the BIC.
Generally, this limiting behavior holds for $g_0=n^k$ for any value $k>0$ (assuming $\delta=n$) with the approximation rate depending on $k$.
For $k>1$, the proof is similar to the one presented above with $k=2$.
For $k=1$, $w=1/2$ and thus the dimensionality penalty becomes equal to $\log(1+n/2)$. Therefore for large $n$ again the
PCEP $g$-prior has the same limiting behavior as the BIC but with a slower convergence rate than before.
Finally for $0<k<1$, the dimensionality penalty will be approximately equal to $\log(1+n^{k}) \approx k \log(n)$ which again
for large values of $n$ will become equivalent to the penalty induced by BIC but with an even slower convergence rate. 
Finally, it is well known \cite{fernandez_etal_2001} that consistency holds for BIC under a minor and realistic assumption; see for example Equation 22 in 
\citep{liang_etal_2008}.

\section{Comparison between the PCEP and the Zellner's $g$-prior}
\label{interpretation}

The structure of PCEP $g$-prior is similar to the structure of the Zellner's $g$-prior but with different covariance matrix, for given $\sigma^2$.
As we will see our prior leads to a variable-selection procedure that it is more parsimonious than the one using Zellner's $g$-prior with $g=n$ taking into account uncertainty of imaginary
data generated from the null model as reference.

We compare theoretically the volumes of the covariance matrices, the maximum prior ordinates, and by this way the
dispersions of the two prior distributions. Additionally, we compared graphically, for simulated scenarios, the orientations of the
two prior distributions and the behavior of the posterior model probabilities for a variety of correlations.

From (\ref{PCEP_prior}), we have that for a given $\sigma^2$, the covariance matrix of the PCEP $g$-prior is given by
$$
\Sigma^{PCEP}_\ell=
\delta
\left\{ \bx_\ell^{*^T}
\left[ w^{-1}  \identst - ( \delta \Lambda_0^{*}  + w \mt{H}_\ell^* )^{-1} \right]
\bx_\ell^* \right\}^{-1}.
$$
The determinant of $\Sigma^{PCEP}_\ell$ is given by
\begin{eqnarray}
|\Sigma^{PCEP}_{\ell}| \hspace{-0.3cm}
& = &  \left[ \delta w (w+1) \right]^{d_\ell - d_0} g_0^{d_0} \big| \mt{X}_\ell^{*T} \mt{X}_\ell^* \big|^{-1} \label{step2};
\end{eqnarray}
detailed derivation of this expression is given at the Appendix \ref{eq_step}.

%


If we set in the PCEP $g$-prior $g_0=n^2$ and $\delta=n$ then the volume variance multiplicator appearing in (\ref{step2}) becomes equal to
$$
\left[ \delta w (w+1) \right]^{d_\ell - d_0} g_0^{d_0} = n^{2 d_\ell} \left[ \frac{ 2n + 1 }{ (n+1)^2 } \right]^{d_\ell-d_0}
$$
which is greater than $n^{d_\ell}$, i.e. the corresponding multiplicator in the Zellner's $g$-prior with $g=n$,
for any sample size $n\ge 2$.
This can be easily proved if we consider the function
$$
\phi(n) = \log \left( n^{d_\ell} \left[ \frac{ 2n + 1 }{ (n+1)^2 } \right]^{d_\ell-d_0}  \right)
$$
which is the logarithm of the ratio of the two multipliers. This is an increasing function of $n$ since
$$
\frac{\partial \phi(n)}{\partial n} =\frac{d_\ell(3n+1)}{n(2n+1)(n+1)}+\frac{2d_0n}{(2n+1)(n+1)} \geq 0
$$
and furthermore $\phi(2) = \log \left[ 2^{d_\ell} \left( \frac{ 5 }{ 9 } \right)^{d_\ell-d_0} \right]
= \log \left[   \left( \frac{ 10 }{ 9 } \right)^{d_\ell } \left( \frac{ 9 }{ 5 } \right)^{d_0 } \right] \ge 0 $.

\begin{figure}[htpb]
\caption{Contour plots of PCEP and Zellner's $g$-priors ($g=n$) for various correlation values $r$ and for $n=30$ and $50$}
\label{contour_plots}
\begin{center}
\centerline{(a) $n=30$}
\psfrag{PCI}[c][c][1.0]{ PCEP $g$-prior }
\psfrag{Zellner's g-prior}[c][c][1.0]{ Zellner's $g$-prior }
\includegraphics[scale=0.45, angle=-90]{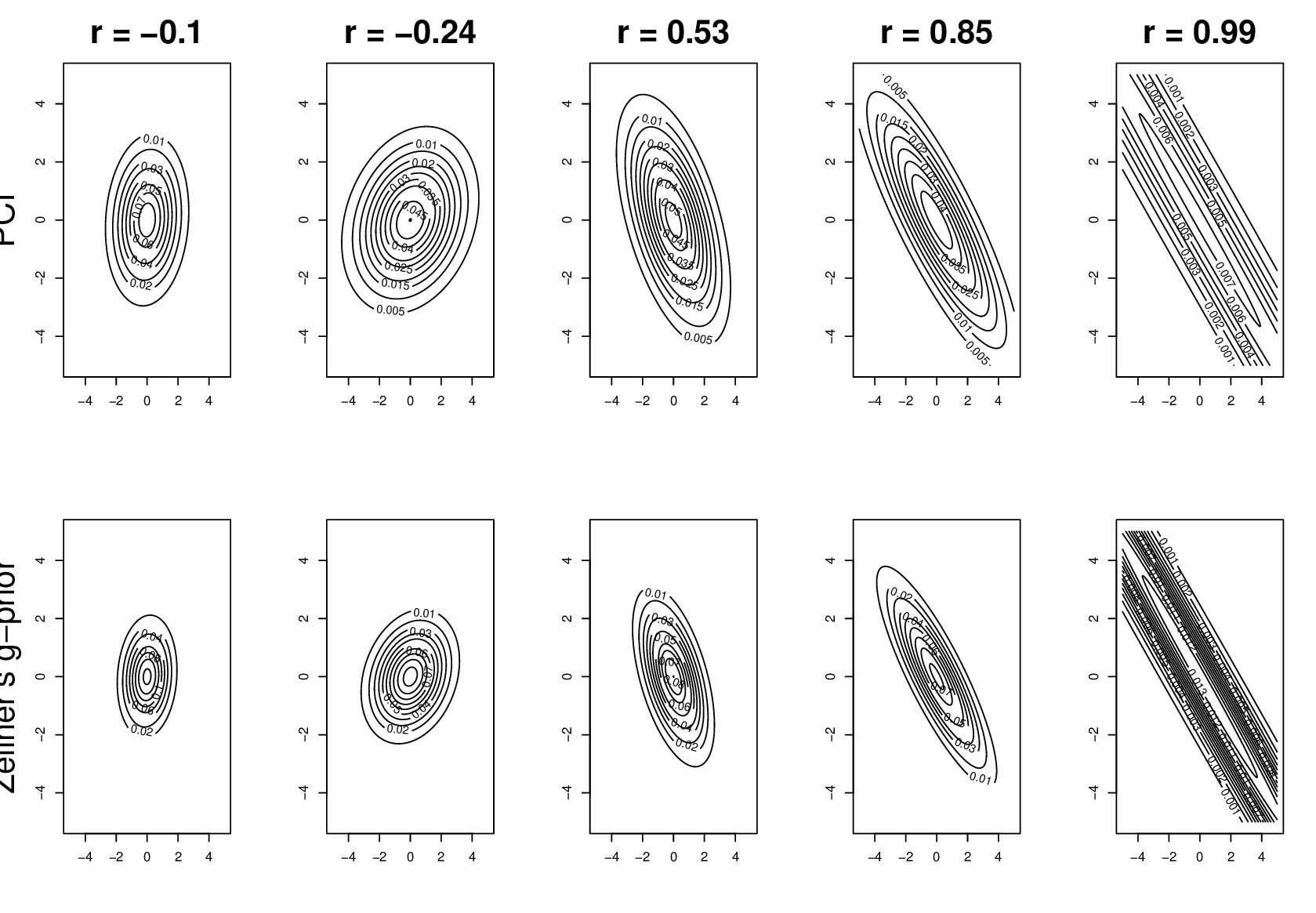}

\centerline{(b) $n=50$}
\psfrag{PCI}[c][c][1.0]{ PCEP $g$-prior }
\psfrag{Zellner's g-prior}[c][c][1.0]{ Zellner's $g$-prior }
\includegraphics[scale=0.45, angle=-90]{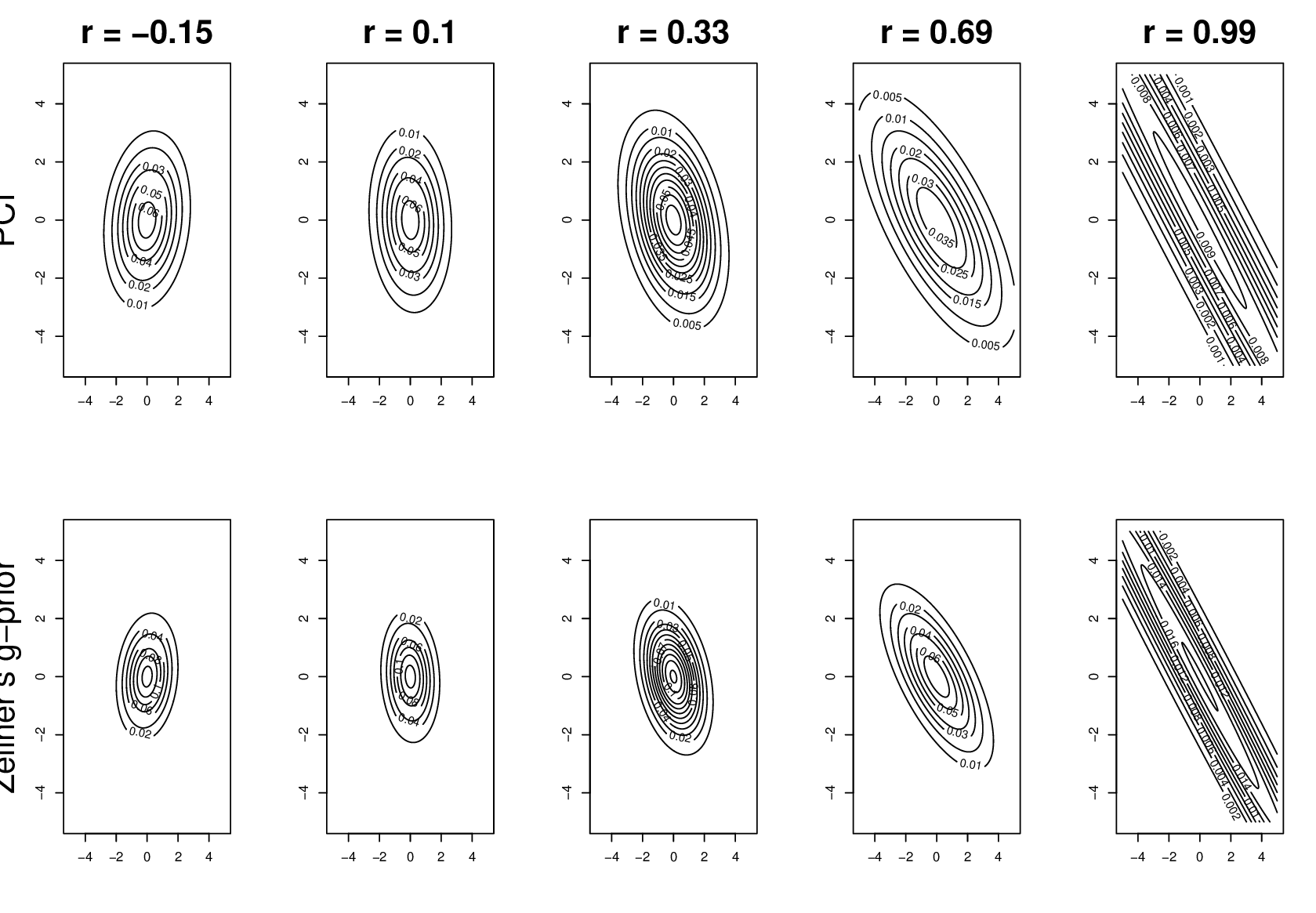}

\end{center}
\end{figure}

For illustration, we have generated two covariates $X_1$ and $X_2$ for a prespecified grid of correlations ($\rho=0,0.1, \dots, 0.9$)
and compared the contour plots of the two prior distributions.
In all cases, PCEP is more dispersed as expected by the above result and the orientation (i.e. prior dependence between $\beta_1$ and $\beta_2$)
remains similar (but not exactly the same) for medium and large sized datasets (e.g. $n\ge 30$).
Some indicative contour plots for various correlation values $r$ and for $n=30$ and $n=50$ are given in Figure \ref{contour_plots}.

Finally, in order to compare the behavior of the posterior model probabilities between the PCEP and the Zellner's $g$-prior
we have created 100 different datasets of $n=100$ observations and $p=2$ covariates.
We have considered different correlation values between the covariates, $Cor(X_1,X_2)=0, 0.1, 0.3, 0.5,0.7, 0.9, 0.99$, while
the response was generated from a model $Y=1+\rho X_1+\sqrt{(1-\rho^2)}\varepsilon$, with $\varepsilon \sim N(0,1)$ and
$\rho=0.2, 0.3, 0.4, 0.5, 0.6$. Under this model, the total variance of $Y$, under each value of $\rho$, is equal to one. 

\begin{figure}[htpb]
\caption{Relative percentage differences (PCEP - Zellner's $g$-prior, $g=n$) between the means and the standard deviations of the posterior probabilities of the true model over 100 generated samples}
\label{image_plots}
\begin{center}

\centerline{(a) Differences between the means }

\includegraphics[scale=0.40]{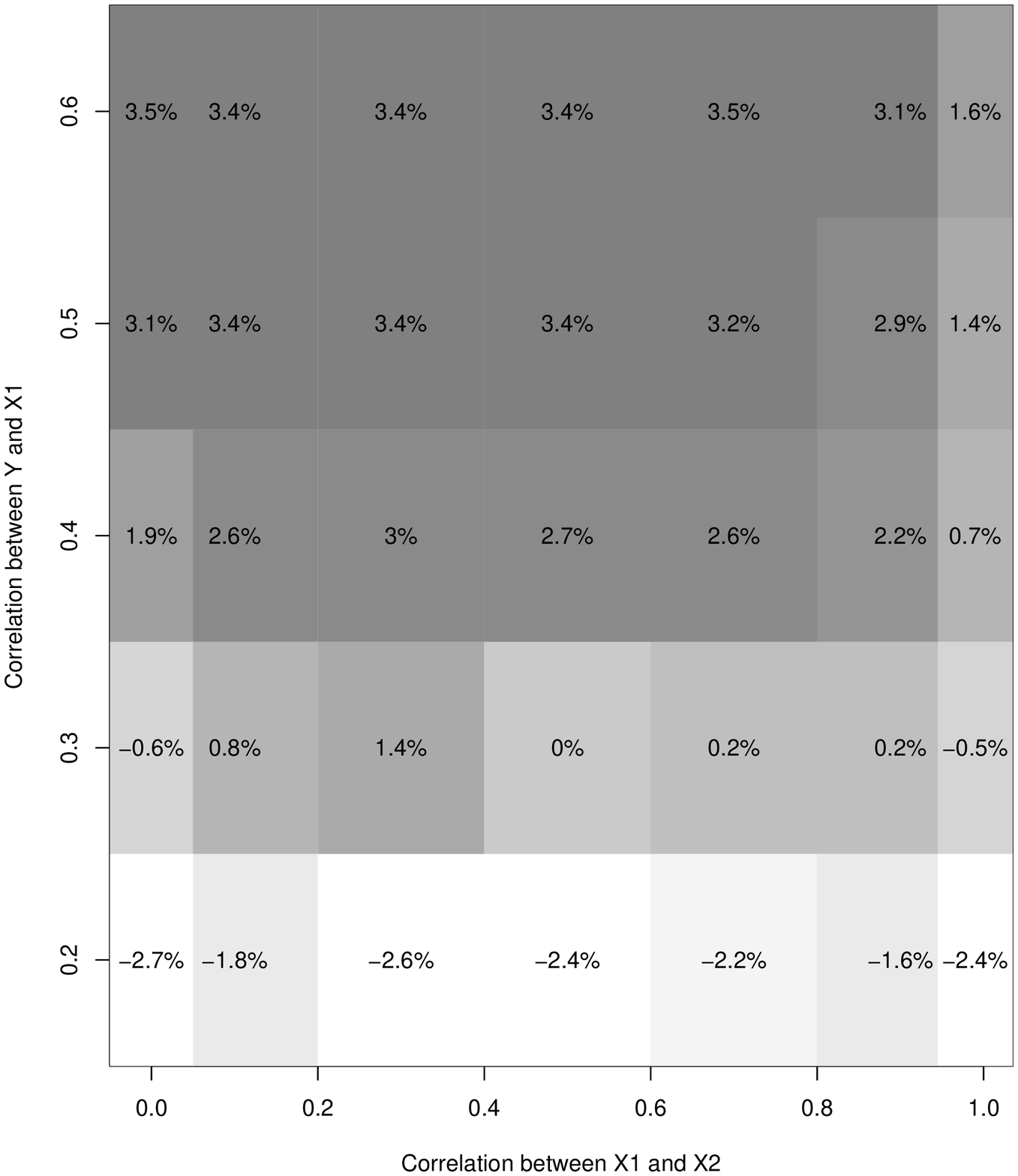}

\centerline{(b) Differences between the standard deviations }
\includegraphics[scale=0.40]{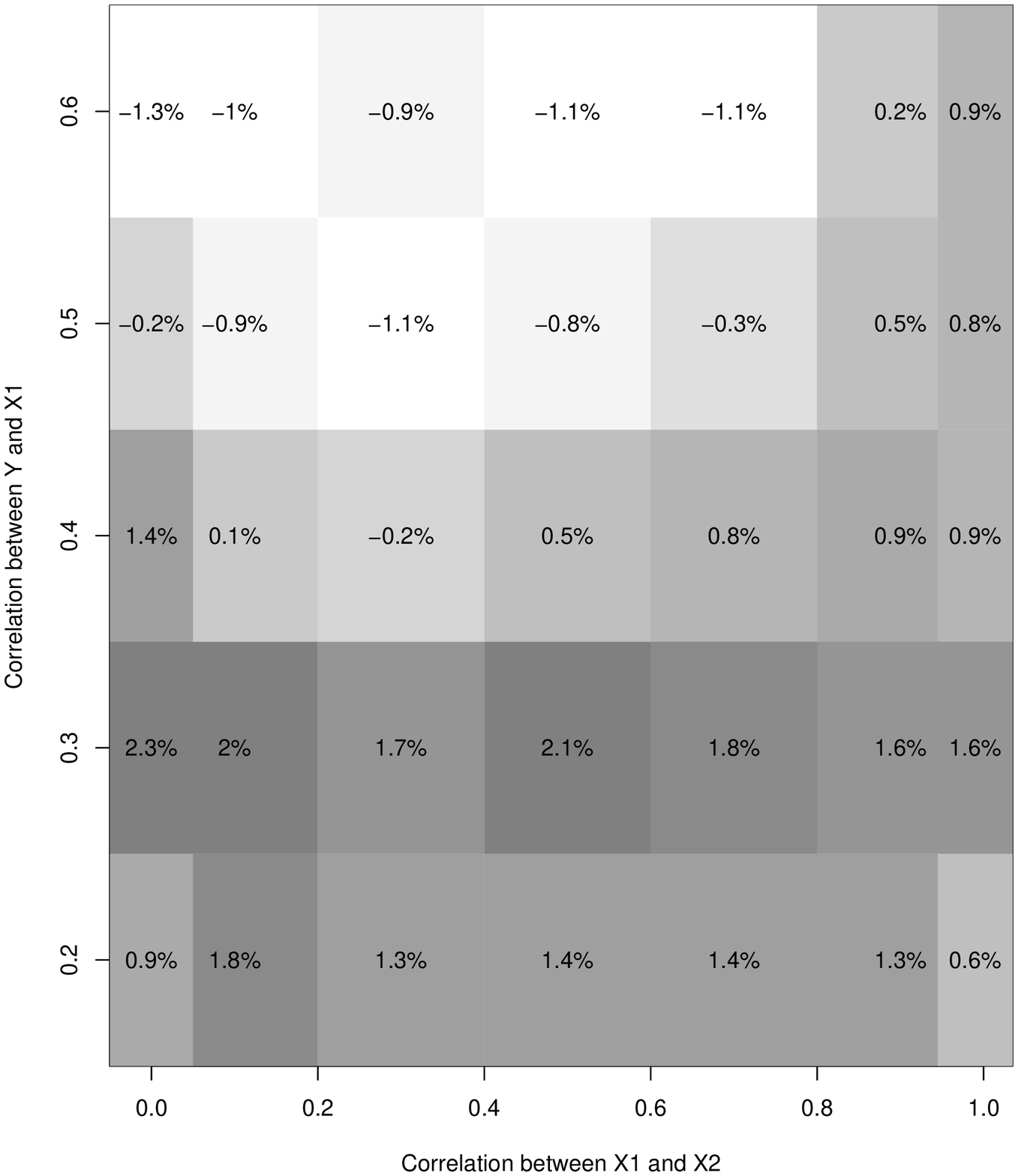}

\end{center}
\end{figure}

Figure \ref{image_plots}(a) presents the relative differences (\%) in the mean posterior model probabilities of the true model over the 100 different generated datasets, while
Figure \ref{image_plots}(b) presents the corresponding differences of the standard deviations. From these two figures it is obvious that PCEP $g$-prior leads to a variable-selection procedure that it is more parsimonious than the one using Zellner's $g$-prior (for the pre-selected hyperparameter values) selecting the true model with a higher weight when the covariates are
higher correlated and with lower weight when the covariates are less correlated. Moreover, the standard deviations of the
posterior model probabilities are higher when using the PCEP $g$-prior, compared to the corresponding ones when using the Zellner's $g$-prior, when
the true effect  of $X_1$ on $Y$ (i.e. $\rho$) is lower. This is a desired property; PCEP $g$-prior gives less posterior weight on average on the
true model when $\rho$ is low but with less certainty compared with Zellner's $g$-prior, while when $\rho$ is large PCEP $g$-prior gives higher
posterior weights on average to the true model with greater confidence.

\section{Experimental results}
\label{examples}

In this section we illustrate the PCEP $g$-prior methodology on both simulated
and real examples. For the implementation of the method we have used the hyperparameter specification described in Section \ref{section4.4}.
We contrast the results of our proposed method using the modified version of the $g$-prior, as in \citep{liang_etal_2008}, with $g=n$, 
the hyper-$g$ prior with $\alpha=3$, as suggested by \citep{liang_etal_2008} and (for the real life example only) the BIC. For the implementation of the $g$-prior 
and the hyper-$g$ prior the \texttt{R} package \texttt{BAS}, available from \texttt{http://www.stat.duke.edu/$\sim$clyde/BAS}, has been used. 

\subsection{Simulation study}

Here we consider the simulated dataset of
\citep{nott_kohn_05}.
This dataset consists of $n=50$ observations and $p=15$ covariates. The first 10 covariates are generated from a standardized normal distribution
while
$$
X_{ij} \sim N\big( ~0.3X_{i1}+0.5X_{i2}+0.7X_{i3}+0.9X_{i4}+1.1X_{i5}, ~1 ~\big) \mbox{~for~} j=11,\dots, 15,~i=1,\dots, 50
$$
and the response from
\begin{eqnarray}
\label{ss}
 Y_i \sim N \big( ~4 + 2X_{i1} - X_{i5} + 1.5 X_{i7} + X_{i11} + 0.5 X_{i13}, ~2.5^2 ~\big), \quad \mbox{for} \quad i=1,\dots,50.
\end{eqnarray}

With $p=15$ covariates we are able to conduct a full enumeration search and avoid extra Monte Carlo variation due to stochastic search of the model space. 

\begin{figure}[h!]
\caption{Boxplots, over 100 random samples, of posterior inclusion probabilities under the three different prior set-ups for the 
simulation study (1: PCEP, 2: $g$-prior with $g=n$, 3: hyper-$g$ prior with $\alpha=3$)}
\label{ex1_incprobs}
\vspace{-3em}
\begin{center}
\includegraphics[scale=0.6, angle=-90]{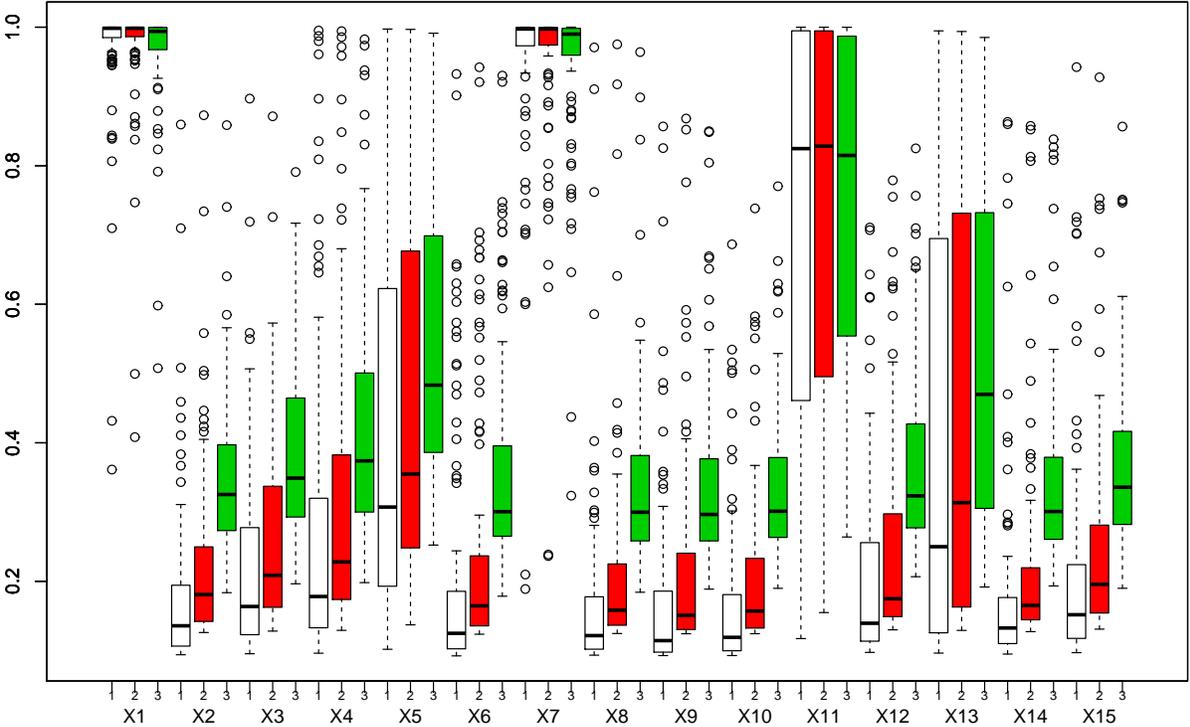}
\end{center}
\vspace{-2em}
\end{figure}

\begin{figure}[h!]
\caption{Density plots of the posterior marginal inclusion probability for $X_{13}$, over the 100 different samples}
\label{ex1_x13}
\vspace{-3em}
\begin{center}
\psfrag{X13}[c][c][0.6]{~$X_{13}$}
\includegraphics[scale=0.6, angle=-90]{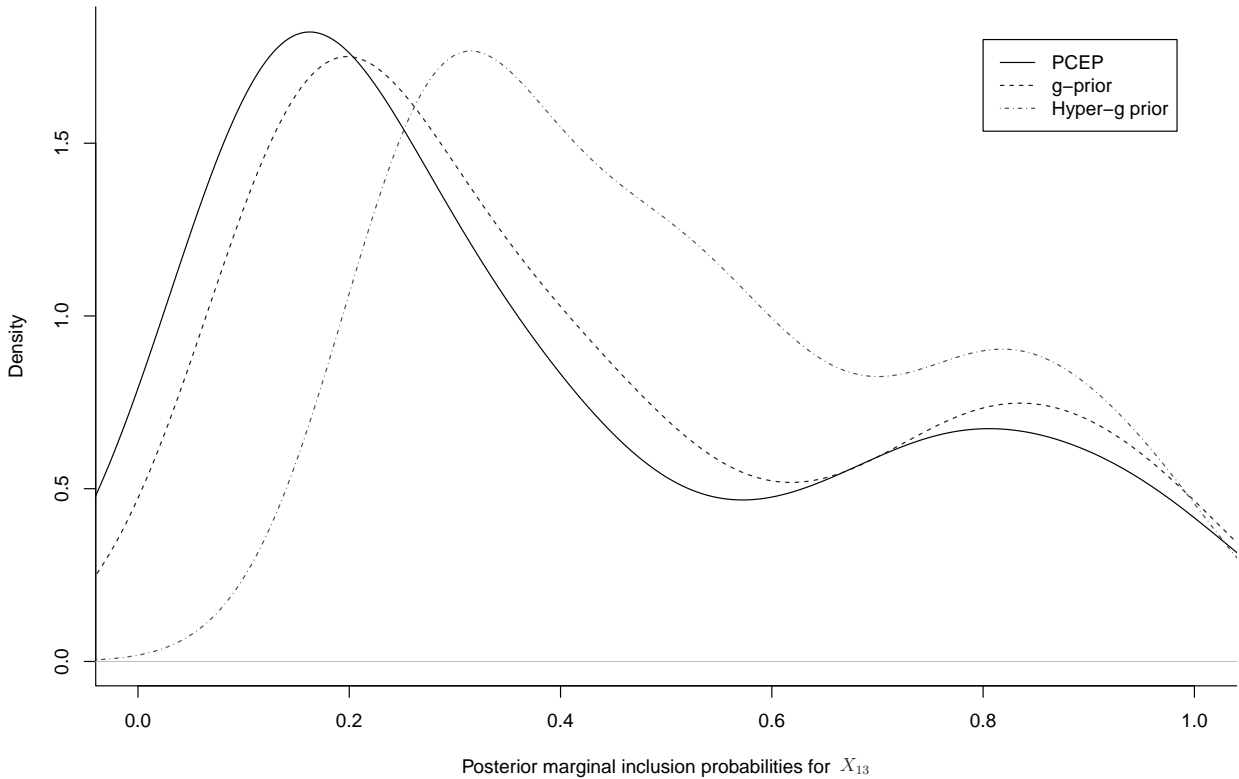}
\end{center}
\vspace{-2em}
\end{figure}

In order to check the efficiency of the proposed method, we generate repeatedly 100 different sets of response variables from the sampling scheme (\ref{ss}). 
Figure \ref{ex1_incprobs} presents boxplots comparing the posterior marginal inclusion probabilities, under 
the three different prior set-ups, over those 100 different samples. No noticeable differences between the boxplots of the posterior marginal inclusion probabilities are observed for the dominating effects of variables $X_1$, $X_7$ and $X_{11}$. For the rest of the 
covariates (i.e. the ones with median posterior marginal inclusion probabilities below 0.5), 
the PCEP based method is systematically more parsimonious, while the hyper-$g$ based procedure supports more complicated models than the other approaches. 
Generally, PCEP shrinks marginal posterior inclusion probabilities towards zero for small effects. 
Figure \ref{ex1_x13} illustrates this behavior; 
it graphically presents the density of the marginal posterior inclusion probability of $X_{13}$ 
over the 100 different samples and under the three different priors. 
Variable  $X_{13}$ was selected due to its large variability of the posterior marginal inclusion probabilities 
under all three prior set-ups, as suggested by Figure \ref{ex1_incprobs}. 
In all three cases, the distribution is bimodal, with the same mode for datasets with clearly non-zero effect 
and a considerably lower mode when using the PCEP prior  for realizations with close to zero estimated effects. 

Similar findings are observed in Table \ref{simtab1} which presents summary statistics of the posterior ranking of the true model. 
All three methods identify the true as the maximum a-posteriori (MAP) model, at least once, but, the PCEP method gives, on 
average, lower rankings to the true model. Additionally, we observe considerably higher variability under the other two prior approaches, with the standard deviation of the ranks under the hyper-$g$ prior 
to be twice as large as the corresponding one under the PCEP prior.

Finally, the hyper-$g$ prior identifies, on average, 3.8 of the 5 non-zero effects, in contrast to the other two procedures that 
identify, on average, 3.4 of the 5 non-zero effects. On the other hand, the PCEP and $g$-prior perform better in regard to the  
identification of the zero effects (9.2 out of 10 in contrast to 8.6 out of 10 for the hyper-$g$ prior).

\begin{table}[h]
\caption{Summary statistics of the posterior ranking of the true model, over 100 repeated samples for the 
simulation study}
\label{simtab1}
\begin{center}
\begin{tabular}{lccccccc}
\hline
Method & Min & $Q_1$  & Median  &  Mean & $Q_3$   &  Max & SD \\
\hline
PCEP & 1.0    & 5.0  & 20.5 &  87.5  & 66.5 & 1733.0 & 226.0 \\
Zellner's $g$-prior ($g=n$) & 1.0  &  4.7 &  22.0 & 110.1  & 95.0 & 2345.0 & 300.5\\
Hyper-$g$ prior ($\alpha=3$) & 1.0   &  4.0  &  29.5  & 232.3 & 207.8 & 4163.0 & 572.0\\
\hline
\multicolumn{6}{p{10cm}}{ \footnotesize $Q_1$ and $Q_3$ denote the first and third quartile respectively} \\ 
\end{tabular}
\end{center}
\end{table}

\subsection{Crime dataset}
\label{crime_data}

Here, we use the crime data \cite{vandaele_1978} to implement the PCEP $g$-prior approach.
\citep{raftery_etal_97} used those data as an illustration of Bayesian model averaging in linear regression using a normal-inverse-gamma
prior for each model parameters, while \citep{fernandez_etal_2001} revisited the crime data exploring the choice of $g$ on $g$-priors.
Finally \citep{liang_etal_2008} used those data for comparing the mixture of $g$ priors formulation with fixed $g$-priors, empirical Bayes approaches and other default procedures.

The data are available in the \texttt{R} package \texttt{MASS} under the name \texttt{UScrime}, and comprise aggregate measures of the crime rate for 47 states
and include 15 explanatory variables. The response variable is the rate of crimes in a particular category per head of population.
All variables, including the response and excluding the indicator covariate ($X_2$), have been
initially log-transformed and then all variables have been centered.

\begin{figure}[h!]
\caption{Posterior inclusion probabilities for the four different prior set-ups for crime data}
\label{ex2_incprobs}
\vspace{-3em}
\begin{center}
\includegraphics[scale=0.6, angle=-90]{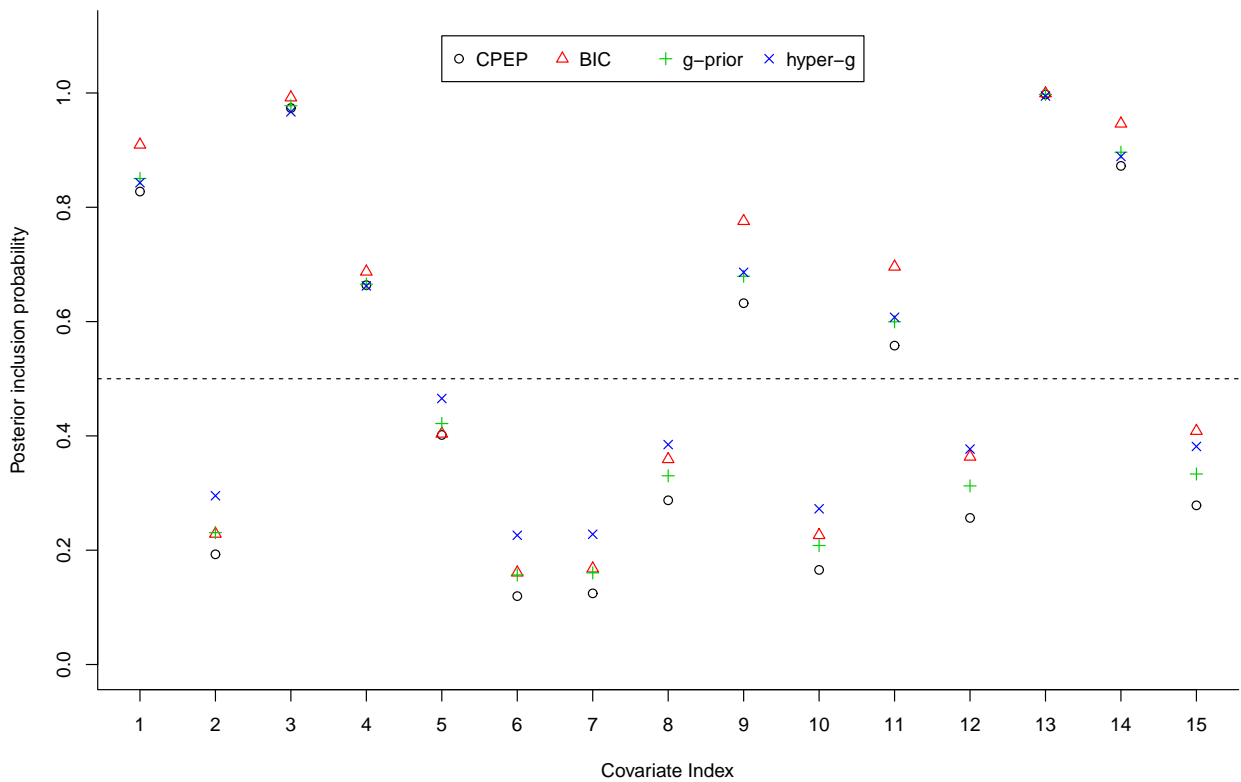}
\end{center}
\vspace{-2em}
\end{figure}

Again, with $p=15$ covariates we were able to contact a full enumeration search. Here additionally to the three different prior set-ups, we present results for BIC. Posterior marginal inclusion probabilities, are presented in Table \ref{crimetab1} and in Figure \ref{ex2_incprobs}. 
We see that all four methods give approximately equal support to the most prominent covariates, while for the remaining ones
the posterior inclusion probabilities are lower under the PCEP approach. Posterior model odds and rankings for the five best PCEP models under all competing approaches  are given in Table \ref{crimetab2}. We notice that all methods support the same two models as the two best ones, model $m_{(1)}$ that includes covariates $X_1, X_3, X_4, X_9, X_{11}, X_{13}, X_{14}$ and model $m_{(2)}$ which is the same as model $m_{(1)}$ with the addition of covariate $X_{15}$.  For all four model selection procedures, the posterior odds of $m_{(1)}$  versus $m_{(2)}$,  range from 0.76 (for BIC) to 1.25 (for PCEP). These differences do not suggest that any of the two models dominate over the other. For the remaining three models no firm conclusion can be drawn except that the third model under PCEP (with only six covariates) is placed in a much lower position under all three remaining approaches. 

\begin{table}[h]
\caption{Posterior marginal inclusion probabilities for the crime data}
\label{crimetab1}
\begin{center}
\begin{tabular}{l@{~~}l c c@{~~} c@{~~} c}
\hline
&                        &      &         &  Zellner's           &  Hyper-$g$           \\
& Variables (log scale)  & PCEP & BIC     &  $g$-prior ($g=n$)   &  prior ($\alpha=3$)  \\
\hline

$X_{1}$ & Percentage of males aged 14-24                & 0.828 &   0.909 & 0.850 & 0.843 \\
$X_{2}$ & Indicator variable for a Southern state       & 0.193 &   0.229 & 0.231 & 0.295 \\
$X_{3}$ & Mean years of schooling                       & 0.974 &   0.992 & 0.978 & 0.967 \\
$X_{4}$ & Police expenditure in 1960                    & 0.664 &   0.687 & 0.665 & 0.662 \\
$X_{5}$ & Police expenditure in 1959                    & 0.402 &   0.404 & 0.422 & 0.465 \\
$X_{6}$ & Labour force participation rate               & 0.120 &   0.161 & 0.157 & 0.226 \\
$X_{7}$ & Number of males per 1000 females              & 0.124 &   0.168 & 0.160 & 0.228 \\
$X_{8}$ & State population                              & 0.287 &   0.359 & 0.330 & 0.385 \\
$X_{9}$ & Number of non-whites per 1000 people          & 0.632 &   0.776 & 0.679 & 0.686 \\
$X_{10}$ & Unemployment rate of urban males 14-24       & 0.165 &   0.226 & 0.208 & 0.272 \\
$X_{11}$ & Unemployment rate of urban males 35-39       & 0.558 &   0.696 & 0.600 & 0.608 \\
$X_{12}$ & Gross domestic product per head              & 0.256 &   0.363 & 0.312 & 0.377 \\
$X_{13}$ & Income inequality                            & 0.997 &   0.999 & 0.997 & 0.995 \\
$X_{14}$ & Probability of imprisonment                  & 0.872 &   0.946 & 0.896 & 0.889 \\
$X_{15}$ & Average time served in state prisons         & 0.278 &   0.409 & 0.333 & 0.382 \\
\hline
\end{tabular}
\end{center}
\end{table}

\begin{table}[h]
\begin{center}

\caption {Posterior odds and rankings for the five best PCEP models under all methods, for the crime data} 

\label{crimetab2}

\vspace*{0.15in}

Common variables in all models: $X_1 + X_3 + X_{13} + X_{14}$

\vspace*{0.15in}

\begin{tabular}{c|c@{}c@{}c@{}c@{}c@{}c@{ }|c@{ }|cccc}

\hline
  &  \multicolumn{6}{c|}{Additional} & Number of & \multicolumn{4}{c}{Posterior odds $PO_{ 1 k }$}\\
  $k$ & \multicolumn{6}{c|}{Variables} & Covariates &  PCEP & BIC & $g$-prior & hyper-$g$ prior\\
\hline \hline
 1  &$+X_4$ &        &$+X_9$ &$+X_{11}$ & &          & 7 &1.00 & (2) 1.00 &~(1) 1.00 &~~(2) 1.00\\
 2  &$+X_4$ &        &$+X_9$ &$+X_{11}$ & &$+X_{15}$ & 8 &1.25 & (1) 0.76 &~(2) 1.03 &~~(1) 0.93\\
 3  &$+X_4$ &        &       &$+X_{11}$ & &$+X_{15}$ & 6 &1.40 & ($>50$)  $>9$&(46) 6.88 &($>50$)  $>5$\\
 4  &       & $+X_5$ &$+X_9$ &$+X_{11}$ & &          & 7 &1.56 & (5) 1.61&(13) 2.87 & ~~(3)  1.45\\
 5  &$+X_4$ &        &$+X_9$ &          & &          & 6 &2.07 &26 4.39 &~(3) 1.52 & ~(18) 2.87\\
\hline
\multicolumn{11}{p{13cm}}{ \footnotesize $PO_{ 1 k }$ denote the posterior odds of the PCEP MAP model versus current model $k$} \\
\end{tabular}

\end{center}

\end{table}

\paragraph{Comparison of the predictive performance.}
Here we examine the out-of-sample predictive performance of PCEP, $g$-prior ($g=n$) and hyper-$g$ prior ($\alpha=3$) on the full model and the
two MAP models indicated by PCEP and hyper-$g$ prior in the previous analysis. To do so, we randomly partitioned
the data in half 50 times. For each partition, in order to measure the predictive performance of each model we compute the root mean square error for the validation dataset $V$ of size $n_V=\big[\tfrac{n+1}{2}\big]$
\begin{equation}
\label{ARMSE}
RMSE_\ell = \sqrt{ \frac{1}{n_V} \sum_{i \in {\cal V}} \big( y_i -  \widehat{y}_{i|m_\ell} \big)^2 };
\end{equation}
here
$\widehat{y}_{i|m_\ell}=\mt{X}_{\ell (i)} \widetilde{\dn{\beta}}_\ell$ is the predicted value
of $y_i$ according to the assumed model $m_\ell\,$,
$\widetilde{\dn{\beta}}_\ell$ is the posterior mean of $\dn{\beta}_\ell$
and
$\mt{X}_{\ell (i)}$ is the $i$-th row of matrix $\mt{X}_\ell$ of model $m_\ell$.

Results for the full model and the two MAP models are given in Table \ref{crimetab5}. For comparison purposes,
we have also included the split-half $RMSE$ measures for these three models using the Zellner's $g$-prior with $g=n$ 
and the mixtures of $g$-prior with $\alpha=3$ as implemented by \texttt{BAS} package in \texttt{R}. In general the differences in 
RMSE are not large enough to infer towards the superiority of the predictive ability of one method. 

\begin{table}[!h]

\caption{Comparison of the predictive performance of the full and the two highest a-posteriori models for the crime data}
\label{crimetab5}
\begin{center}
\begin{tabular}{l@{~~}  c c  c  r@{}l  r@{}l  r@{}l@{}}
\hline
                      &          &       &            & \multicolumn{6}{c}{RMSE$^*$} \\ \cline{5-10}
Model                 & $d_\ell$ &$R^2$  & $R^2_{adj}$& \multicolumn{2}{c}{PCEP} & \multicolumn{2}{c}{$g$-prior} & \multicolumn{2}{c}{hyper-$g$ prior} \vspace{0.25em}\\
\hline
PCEP MAP          &  7 & 0.8268 & 0.7973 & 0.2262 & (0.0346)& 0.2264 & (0.0347) & 0.2262  & (0.0329)\\
hyper-$g$ MAP     &  8 & 0.8420 & 0.8087 & 0.2320 & (0.0387)& 0.2322 & (0.0387) & 0.2310  & (0.0381)\\
full              &  15 & 0.8685 & 0.8064 & 0.3133 & (0.0695)& 0.3136 & (0.0697) & 0.2967  & (0.0571)\\
\hline
\end{tabular}
\\[1ex]
\centerline{\footnotesize \it $^*$Mean (standard deviation) over 50 different split-half out-of-sample evaluations}
\end{center}
\end{table}

\section{Discussion}
\label{sec_discussion}

In this article we explore how random imaginary data can be used to extend the $g$-prior which is a popular default choice in Bayesian variable selection. We link approaches based on priors traditionally used in the objective Bayesian variable selection field, such as the intrinsic \cite{casella_moreno_2006}, the 
expected-posterior  \cite{perez_berger_2002} and the more recent power-expected-posterior \cite{fouskakis_et.al_2013} priors,
with the most dominant Bayesian variable selection approaches based on the $g$-prior \cite{zellner_86} 
and its recent extension using mixtures of $g$-priors \cite{liang_etal_2008}. In contrast to our proposed power-conditional-expected-posterior (PCEP) prior, both the $g$-prior and the hyper-$g$ prior can be derived assuming fixed imaginary data.
 
We focus on the use of random imaginary data through the introduction of a power-expected-posterior prior 
conditionally on the error variance parameter within the normal linear model formulation. 
The induced prior is a conjugate normal-inverse-gamma prior, resulting in a variable selection procedure with similar large sample 
properties to BIC, supporting more parsimonious models than the approach using $g$-prior or hyper-$g$ prior in finite samples. The BIC assymptotic behaviour 
of the PCEP Bayes factors ensures consistency of the PCEP model selection approach.

Future extensions of this approach includes the introduction of an additional hyperprior on the 
power parameter, which plays a similar role as the ``$g$'' parameter under the $g$-prior approach. 
We expect that this approach will retain its approximate BIC behaviour and still being more parsimonious 
than the corresponding procedure using hyper-$g$ prior due to the additional uncertainty introduced by the random imaginary data.

\bibliographystyle{agsm}

\bibliography{biblio3}

\clearpage

\newpage

\pagenumbering{roman}

\appendix

\small

\section*{Appendix to ``Power-Conditional-Expected Priors: Using $g$-priors with Random Imaginary Data for Variable Selection'', by D. Fouskakis and I. Ntzoufras}

\numberwithin{equation}{section}

\section{Derivation of the marginal likelihood $m_\ell^N( \by^* | \sigma^2;\, \bx_\ell^*\, ,\delta )$}

\label{proof_prior_predictive_cond}

\small
\begin{proof}

The marginal likelihood $m_\ell^N( \by^* | \sigma^2;\, \bx_\ell^*\, ,\delta )$
under the baseline prior (\ref{prior_cond}) and the power-likelihood (\ref{pl}), conditional on $\sigma^2$
is given by
\begin{eqnarray*}
m_\ell^N( \by^* | \sigma^2;\, \bx_\ell^*\, ,\delta )
&=& \int f(  \by^* | \dn{\beta}_\ell\,, \sigma^2, m_\ell\,;\bx_\ell^*\,,\delta) \pi_\ell^N \left( \dn{\beta}_\ell\, | \sigma^2 ; \bx_\ell^*\right) d\dn{\beta}_\ell \\
&=& \int f_{N_{n^*}}(  \by^* \,;\, \bx_\ell^* \, \dn{\beta}_\ell\,,
\delta \sigma^2 \mt{I}_{n^*}  )
         f_{N_{d_\ell}} \big( \dn{\beta}_\ell \,;\, \dn{0}, g_0 (\bx_\ell^{*^T} \bx_\ell^*)^{-1} \sigma^2 \big) d\dn{\beta}_\ell \\
&=& \int f_{N_{n^*}}(  \by^* \,;\, \bx_\ell^* \, \dn{\beta}_\ell\,,
\delta\sigma^2 \mt{I}_{n^*}   )
         f_{N_{d_\ell}}\big( \dn{\beta}_\ell \,;\, \dn{0}, \frac{g_0}{\delta} (\bx_\ell^{*^T} \bx_\ell^*)^{-1} \delta \sigma^2 \big) d\dn{\beta}_\ell\,.
\end{eqnarray*}
From the above we have the expression of the marginal likelihood of the 
usual Gaussian regression model with known error variance $\delta\sigma^2$
and a normal conjugate prior with mean zero and variance equal to
$ g_0 \delta^{-1} (\bx_\ell^{*^T} \bx_\ell^*)^{-1} \delta \sigma^2$.
Thus, the marginal likelihood is given by
\begin{eqnarray*}
m_\ell^N( \by^* | \sigma^2;\, \bx_\ell^*\, ,\delta )
&=&  f_{N_{n^*}}\big(  \by^* \,;\, \dn{0}, (\mt{I}_{n^*} + \tfrac{g_0}{\delta} \mt{X}_\ell^{*}(\mt{X}_\ell^{*^T}\mt{X}_\ell^*)^{-1}\mt{X}_\ell^{*^T})  \delta \sigma^2 \big) \\
&=&  f_{N_{n^*}}\big(  \by^* \,;\, \dn{0}, \delta (\mt{I}_{n^*} + \tfrac{g_0}{\delta} \mt{X}_\ell^{*}(\mt{X}_\ell^{*^T}\mt{X}_\ell^*)^{-1}\mt{X}_\ell^{*^T})  \sigma^2 \big) \\
&=&  f_{N_{n^*}}\big(  \by^* \,;\, \dn{0}, {\Lambda_\ell^*}^{-1} \sigma^2
\big)
\end{eqnarray*}
with
\begin{eqnarray*}
{\Lambda_\ell^*}^{-1} &=& \delta \Big(\mt{I}_{n^*} + \tfrac{g_0}{\delta}
\mt{X}_\ell^{*}(\mt{X}_\ell^{*^T}\mt{X}_\ell^*)^{-1}\mt{X}_\ell^{*^T} \Big)
=  \delta\mt{I}_{n^*} +  g_0  \mt{X}_\ell^{*}(\mt{X}_\ell^{*^T}\mt{X}_\ell^*)^{-1}\mt{X}_\ell^{*^T}  \Leftrightarrow \\
\Lambda_\ell^* &=& \delta^{-1} \Big(\mt{I}_{n^*} + \tfrac{g_0}{\delta}
\mt{X}_\ell^{*}(\mt{X}_\ell^{*^T}\mt{X}_\ell^*)^{-1}\mt{X}_\ell^{*^T}
\Big)^{-1} = \delta^{-1} \Big( \mt{I}_{n^*} -
\tfrac{g_0/\delta}{g_0/\delta +1} \mt{H}_\ell^* \Big)
= \delta^{-1} \big(\mt{I}_{n^*} - \tfrac{g_0}{g_0 + \delta} \mt{H}_\ell^* \big) \\
&=& \delta^{-1} \big(\mt{I}_{n^*} - w \mt{H}_\ell^* \big)
\end{eqnarray*}
and $\mt{H}_\ell^*= \mt{X}_\ell^{*}(\mt{X}_\ell^{*^T}\mt{X}_\ell^*)^{-1}\mt{X}_\ell^{*^T}$.

\end{proof}

\section{Derivation of the conditional posterior $\pi_\ell^N(\dn{\beta}_\ell\, | \by^*, \sigma^2\,;\, \bx_\ell^*\, ,\delta)$}

\label{proof_cond_posterior}
\begin{proof}

\begin{eqnarray*}
\pi_\ell^N(\dn{\beta}_\ell\, | \by^*, \sigma^2\,;\, \bx_\ell^*\, ,\delta) &=&
\frac{f(\by^* \, | \dn{\beta}_\ell\, , \sigma^2, m_\ell\; ; \bx_\ell^*\, ,\delta)
        \pi_\ell^N \left( \dn{\beta}_\ell\, | \sigma^2 ; \bx_\ell^*\right)}
     {m_\ell^N (\by^* | \sigma^2;\,\bx_\ell^*\, ,\delta)}  \\
&\propto&
f(\by^* \, | \dn{\beta}_\ell\,, \sigma^2, m_\ell \,;\bx_\ell^*\,,\delta) \pi_\ell^N \left( \dn{\beta}_\ell\ | \sigma^2 ; \bx_\ell^* \right) \\
&\propto& f_{N_{n^*}}(\by^* \,;\, \mt{X}_\ell^* \dn{\beta}_\ell\,,
\delta \sigma^2 \mt{I}_{n^*}  )
                f_{N_{d_\ell}} \left(\dn{\beta}_\ell\,;\, \dn{0},g_0( \mt{X}_\ell^{*^T} \mt{X}_\ell^*)^{-1} \sigma^2\right) \\
&\propto& f_{N_{n^*}}(\by^* \,;\, \mt{X}_\ell^* \dn{\beta}_\ell\,,
\delta \sigma^2 \mt{I}_{n^*}  )
                f_{N_{d_\ell}} \left(\dn{\beta}_\ell\,;\, \dn{0},\frac{g_0}{\delta} ( \mt{X}_\ell^{*^T} \mt{X}_\ell^*)^{-1} \delta \sigma^2 \right) \\
& = & f_{N_{d_\ell}} \left(\dn{\beta}_\ell\,;\, w \widehat{\dn{\beta}}_\ell^*, w( \mt{X}_\ell^{*^T} \mt{X}_\ell^*)^{-1} \delta \sigma^2  \right) \\
& = & f_{N_{d_\ell}} \left(\dn{\beta}_\ell\,;\, w
\widehat{\dn{\beta}}_\ell^*, \delta  w( \mt{X}_\ell^{*^T}
\mt{X}_\ell^*)^{-1} \sigma^2  \right),
\end{eqnarray*}
where $w = \dfrac{ g_0/\delta } { g_0/\delta +1 } = \dfrac{ g_0 } { g_0
+\delta } $.
\end{proof}

\section{Model search algorithm}
\label{sec_model_search}

For any number of variables $p$ under consideration in our model uncertainty problem,
the number of models for which we need to evaluate the marginal likelihood is equal to $2^p$.
Reasonably when $p$ is even moderately large, the number of models under consideration
grows tremendously.
As a result, full enumeration of the marginal likelihoods and the corresponding
posterior model weights needed in Bayesian variable selection and evaluation problems
becomes infeasible.
For this reason, in such problems, advanced MCMC methods are used, as
model search algorithms, to trace the most important models and variables.
Estimation of posterior model weights and posterior model odds can be then
made efficiently within reduced model spaces in which unimportant variables
have been excluded according to our model search algorithm;
see \citep{fouskakis_etal_2009} for an example of such practice.

When the marginal likelihood is given in a closed form,
we may use the Markov chain Monte Carlo model composition
\citea{madigan_york_95}{$MC^3$,}.
Posterior model weights can be estimated by both considering
the marginal likelihoods of the visited and proposed  models stored in step 2 (of the algorithm presented below) or
by a simple frequency tabulation of the visited models given by the output of the MCMC sampler.

Under the PCEP approach,
the marginal likelihood is analytically given by expression (\ref{marginal_likelihood_power_cond_intrinsic}).
Hence $MC^3$ can be directly used to explored the model space.
Here we consider the following modified approach of the $MC^3$
in which we sample the binary vector $\dn{\gamma}$, indicating the variables included in the model \citea{george_mcculloch_93}{see for example},
using a Metropolis within Gibbs approach.

\begin{enumerate}
\item For the current model $m_\ell\,$, corresponding to the set of variable inclusion indicators $\dn{\gamma}_\ell$ repeat the following:
            \begin{itemize}
            \item[~]  For $j = 1, \dots, p$ (selected in random order) repeat the following steps:
                    \begin{enumerate}
                    \item Propose $\gamma_j' = 1 - \gamma_j$ with probability equal to one.
                    \item Set the remaining covariates the same i.e. $\gamma_l' = \gamma_l$ for all $l\ne j$.
                    \item Identify $m_{\ell\,'}$ that corresponds to the vector $\dn{\gamma}_{\ell\;'}$ with elements $\gamma_k'\,, k=1, \dots, p$.
                    \item If $m_{\ell\,'}$ is not previously visited,
                                calculate and store its marginal likelihood \newline
                                $m^{PCEP}_{\ell\,'}( \by | \mt{X}_{\ell\,'}\,,\mt{X}_{\ell\,'}^*\, ,\delta)$
                                given by (\ref{marginal_likelihood_power_cond_intrinsic}).
                    \item Set $m_\ell=m_{\ell\,'}$ (i.e. accept the proposed model $m_{\ell\,'}$) with probability
                    \be
                    \alpha = \min \left(  1, \frac{  \pi(m_{\ell\,'}|\by) } {  \pi(m_\ell|\by) }  \right)
                    \equiv \min \left(  1, \frac{  m^{PCEP}_{\ell\,'}( \by | \mt{X}_{\ell\,'}\,,\mt{X}_{\ell\,'}^*\, ,\delta)}
                                           {  m^{PCEP}_\ell( \by | \mt{X}_\ell\,,\mt{X}_\ell^*\, ,\delta)}
                                           \times \frac{ \pi(m_{\ell\,'}) }{ \pi(m_{\ell}) }
                                           \right)
                    \label{acc_prob_mc3_cpi}
										\ee
										where $\pi(m_{\ell})$ is the prior probability of model $m_{\ell}$.
                    \end{enumerate}
                    \end{itemize}
\item Store $m_\ell$ as the current model.
\item Repeat steps 1--2 until a sufficient number of models is visited.
 \end{enumerate}

\section{Derivations for Section \ref{consistency}}
\label{appendix_calc_concistency}

\subsection*{Derivation of equation \ref{consistency1}}

The matrix determinant Lemma \cite[p. 416]{harville_97} states that
\begin{equation}
\label{mp}
|\mt{A}+\mt{C} \mt{B} \mt{D}^T | = |\mt{B}| ~|\mt{A}| ~|\mt{B}^{-1} + \mt{D}^T\mt{A}^{-1} \mt{C} |
\end{equation}
for any $\mt{A}$ and $\mt{B}$ square invertible matrices.
Therefore, we have that
\begin{equation}
\big| \mt{I}_n + \mt{X}_\ell \mt{V_{\dn{\beta}_\ell}} \mt{X}_\ell^T \big|
=  \big| \mt{V_{\dn{\beta}_\ell}} \big|  \big| \mt{V_{\dn{\beta}_\ell}}^{-1} + \mt{X}_\ell  \mt{X}_\ell^T \big|.
 \label{cons1_eq1}
\end{equation}
Similarly from (\ref{v_star}) and (\ref{mp}), we have that
\begin{eqnarray*}
\big| \mt{V_{\dn{\beta}_\ell}^*} \big|
&=& \delta^{\, d_\ell}
\left| w^{-1} \bx_\ell^{^T}\bx_\ell - \bx_\ell^{^T}( \delta \Lambda_0 + w \mt{H}_\ell )^{-1}\bx_\ell \right|^{-1}\\
&=& \delta^{\, d_\ell} w^{d_\ell} \big|  \bx_\ell^{T}\bx_\ell \big|^{-1}
\left| \delta \Lambda_0 \right|^{-1} \left|  \delta \Lambda_0 + w \mt{H}_\ell \right|
\end{eqnarray*}
with $\mt{H}_\ell = \mt{X}_\ell(\mt{X}_\ell^{T}\mt{X}_\ell)^{-1}\mt{X}_\ell^{T}$ and
\begin{eqnarray*}
\big| \mt{V_{\dn{\beta}_\ell}}^{ -1} + \mt{X}_\ell  \mt{X}_\ell^T \big|
&=&
\big| \delta^{-1} w^{-1} \bx_\ell^{T}\bx_\ell - \delta^{-1}\bx_\ell^{T}( \delta \Lambda_0 + w \mt{H}_\ell )^{-1}\bx_\ell
+ \mt{X}_\ell  \mt{X}_\ell^T \big| \\
&=&
\big| \frac{1 + \delta w }{ \delta w } \bx_\ell^{T}\bx_\ell - \delta^{-1} \bx_\ell^{T}( \delta \Lambda_0 + w \mt{H}_\ell )^{-1}\bx_\ell \big| \\
&=&
\left( \frac{ 1 + \delta w }{ \delta w } \right)^{d_\ell} \big|  \bx_\ell^{T}\bx_\ell \big| ~
\big| \delta \Lambda_0 + w \mt{H}_\ell \big|^{-1}
\left|  \delta \Lambda_0 +  \left( \frac{  \delta w^2  }{ 1+\delta w }\right) \mt{H}_\ell   \right|.
\end{eqnarray*}

Returning back to (\ref{cons1_eq1}), we obtain
\begin{eqnarray*}
\big| \mt{I}_n + \mt{X}_\ell \mt{V_{\dn{\beta}_\ell}} \mt{X}_\ell^T \big|
&=&
\delta^{\, d_\ell} w^{d_\ell} \big|  \bx_\ell^{T}\bx_\ell \big|^{-1}
\left| \delta \Lambda_0 \right|^{-1} \left|  \delta \Lambda_0 + w \mt{H}_\ell \right| \\
&&
\times
\left( \frac{ 1 + \delta w }{ \delta w } \right)^{d_\ell} \big|  \bx_\ell^{T}\bx_\ell \big| ~
\big| \delta \Lambda_0 + w \mt{H}_\ell \big|^{-1}
\left|  \delta \Lambda_0 +  \left( \frac{   \delta w^2  }{ 1+\delta w }\right) \mt{H}_\ell   \right| \\
&=&
\left( 1 + \delta w \right)^{d_\ell}
\left|   \Lambda_0 \right|^{-1}
\left|    \Lambda_0 +  \left( \frac{     w^2  }{ 1+\delta w }\right) \mt{H}_\ell   \right|.
\end{eqnarray*}

Since $w=g_0/(g_0+\delta)$, for $g_0>>\delta$, $w\approx 1$ and
$\big| \mt{I}_n + \mt{X}_\ell \mt{V_{\dn{\beta}_\ell}} \mt{X}_\ell^T \big|
=
\left( 1 + \delta \right)^{d_\ell}
\left| \Lambda_0 \right|^{-1}
\left|  \Lambda_0 +  \left( \frac{ 1 }{ 1+\delta  }\right) \mt{H}_\ell   \right|$
which becomes approximately equal to $\left( 1 + \delta \right)^{d_\ell}$ for reasonably large values of $\delta$.

\subsection*{Derivation of equation \ref{consistency2}}
We have that
\begin{eqnarray*}
\by^T \big( \mt{I}_n + \mt{X}_\ell \mt{V_{\dn{\beta}_\ell}} \mt{X}_\ell^T \big)^{-1} \by
&=&
\by^T \Big( \mt{I}_n - \mt{X}_\ell \big( \mt{V}_{\dn{\beta}_\ell}^{-1} + \mt{X}_\ell^T\mt{X}_\ell\big)^{-1}  \mt{X}_\ell^T \Big)  \by \\
&=&
  \by^T\by - \by^T\mt{X}_\ell \big(
  \delta^{-1}
  \left\{ \bx_\ell^{T} \left[ w^{-1} \identst - ( \delta \Lambda_0 + w \mt{H}_\ell )^{-1} \right] \bx_\ell \right\}
   + \mt{X}_\ell^T\mt{X}_\ell\big)^{-1}  \mt{X}_\ell^T \by \\
&=&
  \by^T\by - \delta  \by^T\mt{X}_\ell \big(
   w^{-1} \bx_\ell^{T}\bx_\ell - \bx_\ell^{T}( \delta \Lambda_0 + w \mt{H}_\ell )^{-1}\bx_\ell
   + \delta \, \mt{X}_\ell^T\mt{X}_\ell\big)^{-1}  \mt{X}_\ell^T \by\\
&=&
  \by^T\by - \delta  \by^T\mt{X}_\ell
  \left(
   \left[ \frac{1+\delta w}{w} \right]  \bx_\ell^{T}\bx_\ell - \bx_\ell^{T}(  [\mt{I}_n - w\mt{H}_0] + w \mt{H}_\ell )^{-1}\bx_\ell
   \right)^{-1}  \mt{X}_\ell^T \by\\
&=&
  \by^T\by - \frac{w \delta }{ 1+w \delta  }  \by^T\mt{X}_\ell
  \left(
   \bx_\ell^{T}\bx_\ell - \frac{ w  }{1+w \delta }
   \bx_\ell^{T}( \mt{I}_{n}  + w [\mt{H}_\ell-\mt{H}_0])^{-1}\bx_\ell\right)^{-1}  \mt{X}_\ell^T \by.
\end{eqnarray*}

For the derivation of the first expression, Woodbury's matrix identity  \cite[p. 423--426]{harville_97} has been used.

\normalsize

\subsection*{Derivation of equation \ref{step2}}
\label{eq_step}

Using (\ref{mp}), we obtain
\begin{eqnarray}
|\Sigma^{PCEP}_{\ell}| \hspace{-0.3cm}
&=& \hspace{-0.3cm} \delta^{d_\ell} \Big\{ ~ \big| w^{-1} \mt{X}_\ell^{*T} \mt{X}_\ell^* \big|
                             ~ \big|(-1)( \delta \Lambda_0^{*}  + w \mt{H}_\ell^*) \big|^{-1}
                             ~ \big| -(\delta \Lambda_0^{*}  + w \mt{H}_\ell^*) + w\mt{X}_\ell^*(\mt{X}_\ell^{*T} \mt{X}_\ell^*)\mt{X}_\ell^{*T}  \big|
                           \Big\}^{-1} \nonumber \\
&=& \hspace{-0.3cm} \delta^{d_\ell} w^{d_\ell} \big| \mt{X}_\ell^{*T} \mt{X}_\ell^* \big|^{-1}
    \frac{ \big| \delta \Lambda_0^{*} + w  \mt{H}_\ell^*  \big| }{ \big| \delta \Lambda_0^{*}  \big| }.  \label{appd3_eq1}
\end{eqnarray}

But $\big| \delta \Lambda_0^{*} + w  \mt{H}_\ell^*  \big|$ can be simplified to
\begin{eqnarray*}
\big| \delta \Lambda_0^{*} + w  \mt{H}_\ell^*  \big|
&=& |\delta \Lambda_0^* | |\mt{X}_\ell^{*T} \mt{X}_\ell^* |^{-1} |\mt{X}_\ell^{*T} \mt{X}_\ell^*  + w \delta^{-1} \mt{X}_\ell^{*T} \Lambda_0^{*-1} \mt{X}_\ell^* |
\end{eqnarray*}
using (\ref{mp}).
Moreover, we have that
\begin{eqnarray}
\big| \delta \Lambda_0^{*} + w  \mt{H}_\ell^*  \big|
&=& |\delta \Lambda_0^* | |\mt{X}_\ell^{*T} \mt{X}_\ell^* |^{-1} \left|\mt{X}_\ell^{*T} \mt{X}_\ell^*  + \frac{w}{\delta}  \mt{X}_\ell^{*T}
\Big( \delta \mt{I}_{n^*} + g_0 \mt{X}_0^{*} (\mt{X}_0^{*T}\mt{X}_0^{*})^{-1} \mt{X}_0^{*T}  \Big) \mt{X}_\ell^* \right| \nonumber \\
&=& |\delta \Lambda_0^* | |\mt{X}_\ell^{*T} \mt{X}_\ell^* |^{-1} \left| (w+1) \mt{X}_\ell^{*T} \mt{X}_\ell^*  + \frac{wg_0}{\delta}  \mt{X}_\ell^{*T}
 \mt{X}_0^{*} (\mt{X}_0^{*T}\mt{X}_0^{*})^{-1} \mt{X}_0^{*T}  \mt{X}_\ell^* \right|  \label{appd3_eq2} \\
&=& |\delta \Lambda_0^* | |\mt{X}_\ell^{*T} \mt{X}_\ell^* |^{-1}
(w+1)^{d_\ell}
\left|  \mt{X}_\ell^{*T} \mt{X}_\ell^* \right|
\left| \mt{X}_0^{*T}\mt{X}_0^{*} \right|^{-1} \nonumber \\
&& \times
\left| \mt{X}_0^{*T}\mt{X}_0^{*} + \frac{wg_0}{\delta (w+1)}  \mt{X}_0^{*T} \mt{X}_\ell^{*}
 (\mt{X}_\ell^{*T}\mt{X}_\ell^{*})^{-1} \mt{X}_\ell^{*T}  \mt{X}_0^* \right| \label{appd3_eq4}
\end{eqnarray}
using again (\ref{mp}) in the last determinant of  (\ref{appd3_eq2}).
If we further consider that
\begin{eqnarray*}
\mt{X}_0^{*T} \mt{X}_\ell^{*}  (\mt{X}_\ell^{*T}\mt{X}_\ell^{*})^{-1} \mt{X}_\ell^{*T}  \mt{X}_0^*
 &=&  \mt{X}_0^{*T}  \mt{H}_\ell^* \mt{X}_0^*
 =  \mt{X}_0^{*T}    \mt{X}_0^*,
 \end{eqnarray*}
 since $\mt{X}_0^T \mt{H}_\ell = \mt{X}_0^T$ for any sub-matrix $\mt{X}_0$ of $\mt{X}_\ell$, equation
\ref{appd3_eq4} becomes
\begin{eqnarray}
\big| \delta \Lambda_0^{*} + w  \mt{H}_\ell^*  \big|
&=& |\delta \Lambda_0^* |(w+1)^{d_\ell}
\left| \mt{X}_0^{*T}\mt{X}_0^{*} \right|^{-1}
\left| \mt{X}_0^{*T}\mt{X}_0^{*} + \frac{wg_0}{\delta (w+1)}  \mt{X}_0^{*T}\mt{X}_0^* \right| \nonumber \\
&=& |\delta \Lambda_0^* |
(w+1)^{d_\ell}
\left| \mt{X}_0^{*T}\mt{X}_0^{*} \right|^{-1}
\left| \frac{ \delta w + \delta + wg_0}{\delta (w+1)} \mt{X}_0^{*T}\mt{X}_0^{*}  \right| \nonumber \\
&=& |\delta \Lambda_0^* |
(w+1)^{d_\ell - d_0}
\left( \frac{ \delta w + \delta + wg_0}{\delta  } \right)^{d_0} \nonumber \\
&=& |\delta \Lambda_0^* |
(w+1)^{d_\ell - d_0}
\left( \frac{ (g_0+\delta) \frac{g_0}{g_0+\delta} + \delta }{\delta  } \right)^{d_0} \nonumber \\
&=& |\delta \Lambda_0^* |
(w+1)^{d_\ell - d_0}
\left( \frac{ g_0 + \delta }{\delta  } \right)^{d_0} \label{appd3_eq5}.
\end{eqnarray}

Substituting (\ref{appd3_eq5}) in (\ref{appd3_eq1}), we have that
\begin{eqnarray*}
|\Sigma^{PCEP}_{\ell}| \hspace{-0.3cm}
&=&
\delta^{d_\ell} w^{d_\ell} \big| \mt{X}_\ell^{*T} \mt{X}_\ell^* \big|^{-1}
\frac{ |\delta \Lambda_0^* | (w+1)^{d_\ell - d_0} \left( \frac{ g_0 + \delta }{\delta  } \right)^{d_0} }
     { \big| \delta \Lambda_0^{*}  \big| } \\
&=&
\delta^{d_\ell} w^{d_\ell}
  (w+1)^{d_\ell - d_0} \left( \frac{ g_0 + \delta }{\delta  } \right)^{d_0}
  \big| \mt{X}_\ell^{*T} \mt{X}_\ell^* \big|^{-1}     \\
&=&
\delta^{d_\ell-d_0} w^{d_\ell}
  (w+1)^{d_\ell - d_0} (g_0 + \delta)^{d_0}
  \big| \mt{X}_\ell^{*T} \mt{X}_\ell^* \big|^{-1}     \\
&=&
[ \delta w (w+1) ]^{d_\ell - d_0} w^{d_0} \, (g_0 + \delta)^{d_0}   \big| \mt{X}_\ell^{*T} \mt{X}_\ell^* \big|^{-1}     \\
&=&
[ \delta w (w+1) ]^{d_\ell - d_0} g_0^{d_0} \, \big| \mt{X}_\ell^{*T} \mt{X}_\ell^* \big|^{-1}.     \\
\end{eqnarray*}

\section*{References in the Appendix}

\renewcommand{\section}[2]{}%

{}

\end{document}